\documentclass[twocolumn]{aastex631}

\usepackage{amsmath} 
\graphicspath{{./}}
\begin{document}

\title{A Uniform Analysis of Gas-phase Metallicity Evolution with $1-3$ Gyr Time Sampling over the Past 12 Billion Years}

\author[0009-0009-1792-7199]{Shweta Jain}
\affiliation{Department of Physics and Astronomy, University of Kentucky, 505 Rose Street, Lexington, KY 40506, USA}\email{shweta.jain@uky.edu}

\author[0000-0003-4792-9119]{Ryan L. Sanders}\affiliation{Department of Physics and Astronomy, University of Kentucky, 505 Rose Street, Lexington, KY 40506, USA}

\author[0000-0002-0101-336X]{Ali Ahmad Khostovan}
\affiliation{Department of Physics and Astronomy, University of Kentucky, 505 Rose Street, Lexington, KY 40506, USA}

\author[0000-0001-5860-3419]{Tucker Jones}\affiliation{Department of Physics and Astronomy, University of California Davis, 1 Shields Avenue, Davis, CA 95616, USA}

\author[0000-0003-3509-4855]{Alice E. Shapley}\affiliation{Department of Physics \& Astronomy, University of California, Los Angeles, 430 Portola Plaza, Los Angeles, CA 90095, USA}

\author[0000-0001-9687-4973]{Naveen A. Reddy}\affiliation{Department of Physics \& Astronomy, University of California, Riverside, 900 University Avenue, Riverside, CA 92521, USA}

\author[0000-0002-8111-9884]{Alex M. Garcia}
\affiliation{Department of Astronomy, University of Virginia,
530 McCormick Road, Charlottesville, VA 22904}
\affiliation{Virginia Institute for Theoretical Astronomy, University of Virginia, Charlottesville, VA 22904, USA}
\affiliation{The NSF-Simons AI Institute for Cosmic Origins, USA}

\author[0000-0002-5653-0786]{Paul Torrey}
\affiliation{Department of Astronomy, University of Virginia, 
530 McCormick Road, 
Charlottesville, VA 22904}
\affiliation{Virginia Institute for Theoretical Astronomy, University of Virginia, Charlottesville, VA 22904, USA}
\affiliation{The NSF-Simons AI Institute for Cosmic Origins, USA}

\author[0000-0002-2583-5894]{Alison Coil}
\affiliation{Department of Astronomy and Astrophysics, University of California, San Diego, La Jolla, CA 92092, USA}

\begin{abstract}
We present a systematic investigation of the evolution of the mass–metallicity relation (MZR) and fundamental metallicity relation (FMR) using uniform metallicity diagnostics across redshifts $z\sim0$ to $z\sim3.3$. We present new Keck/DEIMOS measurements of the [OII]$\lambda\lambda3726,3729$ emission line doublet for star-forming galaxies at $z\sim1.5$ with existing measurements of redder rest-optical lines from the MOSDEF survey. These new observations enable uniform estimation of the gas-phase oxygen abundance using ratios of the [OII], H$\beta$, and [OIII] lines for mass-binned samples of star-forming galaxies in 6 redshift bins, employing strong-line calibrations that account for the distinct interstellar medium ionization conditions at $z<1$ and $z>1$. We find that the low-mass power law slope of the MZR remains constant over this redshift range with a value of $\gamma=0.28\pm0.01$, implying the outflow metal loading factor ($\zeta_\text{out}=\frac{Z_{\text{out}}}{Z_{\text{ISM}}}\frac{\dot{M}_{\text{out}}}{\text{SFR}}$) scales approximately as $\rm \zeta_{out}\propto M_*^{-0.3}$ out to at least $z\sim3.3$. The normalization of the MZR at $10^{10}\ \text{M}_\odot$ decreases with increasing redshift at a rate of $d\log(\text{O/H})/dz =-0.11\pm0.01$ across the full redshift range. We find that any evolution of the FMR is smaller than 0.1~dex out to $z\sim3.3$. We compare to cosmological galaxy formation simulations, and find that IllustrisTNG matches our measured combination of a nearly-invariant MZR slope, rate of MZR normalization decrease, and constant or very weakly evolving FMR. This work provides the most detailed view of MZR and FMR evolution from the present day through Cosmic Noon with a fine time sampling of $1-3$~Gyr, setting a robust baseline for metallicity evolution studies at $z>4$ with {\it JWST}.
\end{abstract}


\section{Introduction}\label{sec:intro}

The interstellar medium (ISM) serves as the cosmic stage where baryonic processes including star formation \citep[e.g.,][]{2012ARA&A..50..531K}, gas accretion from the intergalactic/circumgalactic medium \citep[IGM/CGM;][]{1985ApJ...290..154L, 2005MNRAS.363....2K, 2006MNRAS.368....2D}, and feedback-driven outflows from supernovae \citep[SNe;][]{1994ApJ...430L.105F, 2020A&ARv..28....2V,2021ApJ...918...13S} and active galactic nuclei \citep[AGN;][]{2005Natur.433..604D, 2018MNRAS.479.5385H} leave their defining imprints. Rest-frame optical emission lines are powerful diagnostics for probing the chemical and ionization states of the ISM. Emission-line strengths reflect fundamental properties of the ISM, including star-formation rate (SFR), dust attenuation, gas-phase chemical abundance, gas density, ionization state, and the nature of the ionizing radiation field \citep[for a review, see][]{2019ARA&A..57..511K}.

One key tracer of the baryon cycle that can be probed using strong emission line ratios is the metallicity of the ISM, represented by the gas-phase oxygen abundance (O/H). The gas-phase metallicity is continuously regulated by gas flows into and out of galaxies \citep{2004ApJ...613..898T, 2006ApJ...644..813E, 2017ARA&A..55..389T,2019A&ARv..27....3M}, and is governed by parameters including the mass and metal loading factors of outflowing winds, star-formation efficiency, gas inflow rate, and the enrichment level of accreted gas \citep{2011MNRAS.416.1354D,2013ApJ...772..119L}. Galaxy populations exhibit a well-established scaling relation between stellar mass ($\text{M}_{\ast}$) and gas-phase metallicity (O/H), known as the mass-metallicity relation \citep[MZR; e.g.,][]{2004ApJ...613..898T, 2006ApJ...644..813E, 2006ApJ...647..970L, 2011ApJ...730..137Z, 2013ApJ...765..140A, 2020MNRAS.491..944C}. The MZR has been shown to persist up to at least $z \sim 4$, evolving toward lower metallicities at a given stellar mass with increasing redshift \citep[e.g.,][]{maiolino2008amaze,2014A&A...563A..58T,2015ApJ...799..138S, 2016ApJ...822...42O, 2021ApJ...914...19S}. Recent observations with the James Webb Space Telescope ({\it JWST}) suggest the MZR exists up to $z \sim 10$ \citep[e.g.,][]{2023ApJS..269...33N, 2023ApJ...950L...1S,2024A&A...684A..75C}. The shape and normalization of the MZR evolve with redshift, with the slope at lower masses determined by how mass and metal loading factors of outflows scale with stellar mass \citep[e.g.,][]{2008MNRAS.385.2181F, 2011MNRAS.417.2962P, 2012MNRAS.421...98D, 2013ApJ...772..119L,2014ApJ...791..130Z, 2015MNRAS.449.3274F}.

ISM metallicity is found to have an additional dependence on galaxy SFR, such that metallicity is inversely correlated with SFR at fixed $\text{M}_{\ast}$. This secondary dependence is thought to reflect the baryon cycle, where recent accretion of low-metallicity gas drives SFR higher while lowering the ISM metallicity and subsequent consumption of gas and production of new metals increases metallicity as SFR drops \citep{2011MNRAS.416.1354D,2013ApJ...772..119L, 2018MNRAS.477L..16T}. The three-parameter relation between stellar mass, metallicity, and SFR is commonly referred to as the Fundamental Metallicity Relation (FMR), owing to its invariance with redshift \citep[e.g.,][]{2008ApJ...672L.107E, 2010MNRAS.408.2115M, 2010A&A...521L..53L, 2013ApJ...765..140A, 2018MNRAS.477L..16T}. The FMR does not appear to evolve out to $z\sim3$, suggesting relatively little redshift evolution in outflow metal loading factors and gas fractions at fixed $\text{M}_{\ast}$ and SFR \citep[e.g.,][]{2019A&A...627A..42C, 2020MNRAS.491..944C, 2021ApJ...914...19S}. The study of how gas-phase metallicity scales with both $\text{M}_{\ast}$ and SFR across a range of redshifts can provide crucial constraints on the baryon cycle at different epochs, offering insight into the processes driving galaxy growth and evolution of the galaxy population.

Existing samples of galaxies with spectroscopic metallicity measurements number in the hundred thousands at $z<1$ \citep{2017MNRAS.465.1384C} and thousands at $z>1$ \citep{2014ApJ...795..165S, 2015ApJS..218...15K, 2014ApJ...792...75Z, 2023ApJS..269...33N}, yet significant challenges persist that have prevented a robust evolutionary analysis of the MZR and FMR. A major issue is the lack of uniformity in the line ratios used to infer metallicities for samples at different redshifts. This problem is primarily due to the difficulty of accessing the full suite of bright rest-optical lines ([OII]$\lambda\lambda$3726,3729, H$\beta$, [OIII]$\lambda\lambda$4959,5007, H$\alpha$, [NII]$\lambda$6585) with ground-based spectrographs as emitted optical wavelengths shift into the observed near-infrared (NIR) with increasing redshift. This redshifting makes it challenging to apply consistent metallicity estimators across all redshifts. All of the rest-optical strong lines from [OII] to [NII] can be accessed with optical spectrographs at $z=0-0.5$ and with NIR spectrographs at $z=2.0-2.6$ where they fall in windows of high atmospheric transmission. However, at $z\sim0.5-2$, the full rest-optical range can only be probed with combined optical and NIR observations. In addition, H$\alpha$ and [NII] are inaccessible from the ground at $z\gtrsim3$ where they redshift beyond the $K$ band, though {\it JWST} has now made it possible to access the full rest-optical range at $z>2.7$. As a result, previous studies often relied on a mix of metallicity indicators to include samples at multiple redshifts, potentially introducing substantial systematic biases \citep[e.g.,][]{maiolino2008amaze,2011ApJ...730..137Z,2014A&A...563A..58T}. Alternatively, some studies only included a few samples for which the same set of emission lines were available leading to sparse redshift sampling \citep[e.g.,][]{2021ApJ...914...19S}.

An additional concern for evolutionary studies is the choice of strong-line calibrations employed. It is now widely recognized that star-forming galaxies at $z\gtrsim1$ exhibit distinct ISM ionization conditions relative to galaxies and HII regions at $z = 0$ \citep[e.g.,][]{2014ApJ...795..165S, 2015ApJ...801...88S, steidel2016reconciling, 2016ApJ...816...23S, 2017ApJ...835...88K, strom2017nebular, 2018ApJ...868..117S, 2019ApJ...881L..35S, sanders2020mosdef, 2020MNRAS.499.1652T, 2021MNRAS.505..903C, 2022ApJ...926...80G}. Among other effects, these evolving ionization conditions lead to an evolution in the translation between rest-optical emission-line ratios and O/H. Consequently, commonly applied metallicity calibrations based on $z \sim 0$ datasets are unlikely to yield accurate metallicity measurements at $z > 1$, leading to systematic biases in high-$z$ metallicity measurements. To obtain a robust view of MZR and FMR evolution, it is important to use strong-line calibrations matched to the ionization conditions at the redshift of each sample, such that a different calibration set must be employed for low- and high-redshift galaxies. Alternatively, one can perform photoionization modeling of the full suite of rest-optical line ratios to determine gas-phase abundance \citep[e.g.,][]{2018ApJ...868..117S}, though degeneracies between metallicity and ionization state are common when modeling the strong lines only. Empirical strong-line calibrations appropriate for the more extreme ionization conditions present at high redshifts have been constructed using low-redshift galaxies that are analogs of $z\sim2$ galaxies in their emission-line properties \citep{2018ApJ...859..175B, 2020ApJ...889..161N, 2021MNRAS.504.1237P, 2025arXiv250210499S}. {\it JWST} spectroscopy of high-redshift sources is starting to provide a large number of detections of faint temperature-sensitive auroral emission lines (e.g., [OIII]$\lambda$4363) required to derive direct-method O/H, promising robust calibrations constructed in-situ at high redshift in the near future \citep{2023ApJS..269...33N,2024arXiv241215435C, 2024ApJ...962...24S, 2024arXiv240907455L, 2025arXiv250210499S,2025arXiv250403839C}.

In this work, we perform a uniform MZR and FMR evolution analysis spanning $z\sim0-4$ that employs the same set of emission lines to infer metallicity at all redshifts and utilizes a different set of calibrations at $z<1$ and $z>1$ to account for evolving ISM ionization conditions. We present new Keck/DEIMOS observations covering [OII]$\lambda\lambda$3726,3729 of $\sim120$ star-forming galaxies at $z=1.3-1.7$ that have existing measurements of H$\beta$, [OIII], H$\alpha$, and [NII] from the MOSDEF survey \citep{2015ApJS..218...15K}, completing full coverage of the rest-optical strong lines for this $z\sim1.5$ sample. We draw from published galaxy surveys with optical spectroscopy at $z<1$ and NIR spectroscopy at $z>2$ to assemble a combined sample spanning $z=0$ to $z\sim3.3$ in 6 redshift bins providing $1-3$~Gyr time sampling over the past 12~Gyr of cosmic history, across a uniform mass range $\log(\text{M}_{\ast}/\text{M}_{\odot}) \sim 9.0-11.0$. We utilize this combined sample to obtain the most detailed picture to-date of MZR and FMR evolution out to $z\sim4$, analyzed in a manner that minimizes systematic metallicity biases.

The layout of the paper is as follows. We describe the observations, measurements, samples, and derived galaxy properties in Section~\ref{sec:Methods}. Section~\ref{sec:Z_calculations} describes the methodology and calibrations adopted to infer metallicities. We report the resulting MZR and FMR and their evolution with redshift in Section~\ref{sec:Results}. We discuss the implications of these results in Section~\ref{sec:Discussions}. Finally, in Section~\ref{sec:Conclusions} we summarize our results and conclusions. Throughout the work, we assume a flat $\Lambda$CDM cosmology with $H_0 = 70.0$\,km\,s$^{-1}$\,Mpc$^{-1}$, $\Omega_{\Lambda} = 0.7$, and $\Omega_{m} = 0.3$. Magnitudes are on the AB system \citep{1983ApJ...266..713O} and rest-frame wavelengths are given in air for emission lines. Unless otherwise stated, all uncertainty ranges are 68$\%$ confidence intervals (1$\sigma$) and the term metallicity refers to the gas-phase oxygen abundance. The line ratios analyzed in this work are defined as follows:
\begin{gather*}
    \text{O}3 \equiv \frac{[\text{O III}]\ \lambda5007}{\text{H}\beta},\\
    \text{O}2 \equiv \frac{[\text{O II}]\ \lambda\lambda3726, 3729}{\text{H}\beta}, \\
    \text{O}32 \equiv \frac{[\text{O III}]\ \lambda5007}{[\text{O II}]\ \lambda\lambda3726, 3729}, \\
    \text{R}23 \equiv \frac{[\text{O III}]\ \lambda\lambda4959, 5007 + [\text{O II}]\ \lambda\lambda3726, 3729}{\text{H}\beta},\\
    \text{N}2 \equiv \frac{[\text{N II}]\ \lambda6585}{\text{H}\alpha}, \\
    \text{O}3\text{N}2 \equiv \left(\frac{[\text{O III}]\ \lambda5007}{\text{H}\beta}\right) \bigg{/}\left(\frac{[\text{N II}]\ \lambda6585}{\text{H}\alpha}\right).
\end{gather*}

\section{DATA AND MEASUREMENTS} \label{sec:Methods}
\subsection{The MOSDEF survey}\label{subsec:MOSDEF_survey}

In this paper, we examine the rest-optical spectra of galaxies at $z\sim1.5$ using data from the Multi-Object Spectrometer Far Infrared Exploration \citep[MOSFIRE; ][]{2012SPIE.8446E....M} Deep Evolution Field (MOSDEF) survey \citep{2015ApJS..218...15K}. The MOSDEF survey acquired rest-optical spectra for $\sim1500$ galaxies in three redshift bins ($1.37 \leq z \leq 1.70$, $2.09 \leq z \leq 2.61$, and $2.95 \leq z \leq 3.80$) determined by the redshift ranges in which strong rest-optical emission lines fall in the NIR atmospheric transmission windows. See \citet{2015ApJS..218...15K} for a detailed description of the MOSDEF survey design and data reduction. We focus on sources in the $1.37 \leq z \leq 1.70$ range. This study incorporates ancillary imaging and photometric datasets available for the MOSDEF targets, including data from the CANDELS \citep{2011ApJS..197...35G,2011ApJS..197...36K} and 3D-\textit{HST} \citep{2014ApJS..214...24S,2016ApJS..225...27M} surveys. The MOSFIRE spectra were analyzed to determine the redshifts and measure fluxes of all covered rest-optical emission lines. For $z\sim1.5$ targets, the most prominent of these lines are H$\beta$ and [OIII]$\lambda\lambda4959,5007$ in the $J$ band; and H$\alpha$ and [NII]$\lambda6585$ in the $H$ band. The bluer rest-optical [OII]$\lambda\lambda3726,3729$ lines are only accessible with MOSFIRE at $1.61 < z < 1.70$ in the $Y$ band, and fall blueward of MOSFIRE's wavelength coverage at $z<1.61$. Consequently, out of $\sim300$ MOSDEF targets at $1.37 \leq z \leq 1.70$, only 55 star-forming galaxies have [OII] coverage, 24 of which display detections at S/N$\ge3$.

\subsection{DEIMOS Spectroscopy}\label{subsec:DEIMOS}

To obtain [OII]$\lambda\lambda$3726,3729 measurements for MOSDEF galaxies at $z=1.37-1.61$, where the lines fall outside MOSFIRE's coverage, we obtained 0.5 nights of DEep Imaging Multi-Object Spectrograph \citep[DEIMOS; ][]{2003SPIE.4841.1657F} observations in the AEGIS field. These new observations, described below, increased the number of MOSDEF $z\sim1.5$ targets with [OII] coverage from $55$ to $138$ star-forming galaxies. 

\subsubsection{Mask Design and Observations}\label{subsec:mask_designs}
Three DEIMOS slitmasks were observed in the AEGIS field on 1 June 2021. Observations were conducted under clear photometric conditions, with a typical median seeing of 0.55 arcseconds. Total integration times were 45~min for one mask and 54~min for each of the other two masks. The total integration time for each mask comprised three equal-length individual exposures. The 1200G grating was used with the OG550 cutoff filter and a central wavelength of 8800~\AA, providing approximate wavelength coverage of 7500-10000~\AA\ at a spectral resolution of $R\sim6000$. Targets were assigned to DEIMOS slitmasks by giving the highest priority to spectroscopically-confirmed MOSDEF sources at $1.37\le z\le1.70$, and filling remaining space with $z\ge2$ MOSDEF sources for which DEIMOS covers the rest-ultraviolet and galaxies from the 3D-\textit{HST} catalog with either photometric or grism redshifts at $z=1.10-1.70$. In addition to science targets, each mask had one slit placed on a star of magnitude $R=16.5-17.0$ that was used for telluric correction and final flux calibration. Slit widths were 1$^{\prime\prime}$ and the position angle of each slit was either matched to the photometric major axis, or else at an angle as close to the photometric major axis as possible within the constraints for mask milling. Due to the large field of view of DEIMOS, the majority of the 163 $z\sim1.5$ MOSDEF sources in AEGIS spanning 7 MOSFIRE pointings can be covered with just 3 DEIMOS pointings. In total, DEIMOS spectra were obtained for 100 $z\sim1.5$ MOSDEF primary targets, 93 $z>2$ MOSDEF filler targets, and 176 $z=1.10-1.70$ 3D-\textit{HST} filler targets.

\subsubsection{Data Reduction}
The DEIMOS spectroscopic data were reduced using the publicly available software package \texttt{PypeIt} \citep{2020JOSS....5.2308P}  version \texttt{1.15.0}. The reduction process followed the standard steps described in the \texttt{PypeIt} DEIMOS documentation for bias subtraction, flat-fielding, wavelength calibration, sky subtraction, and coaddition of individual exposures to produce reduced 2D spectra. A custom Python routine was used to perform optimal extractions \citep{1986PASP...98..609H}, generating 1D science and error spectra for each target. The extracted spectra were then flux-calibrated using a sensitivity function derived from the observation of the spectroscopic standard star Feige~34 on the same night as the science observations. The science spectra on each mask were also corrected for telluric absorption by applying the best-fitting telluric model derived from the spectrum of the slit star on that mask.

\subsubsection{Final Flux Calibration and Slit-loss Correction}\label{subsec:sliloss}
\begin{figure*}
\includegraphics[width=\textwidth]{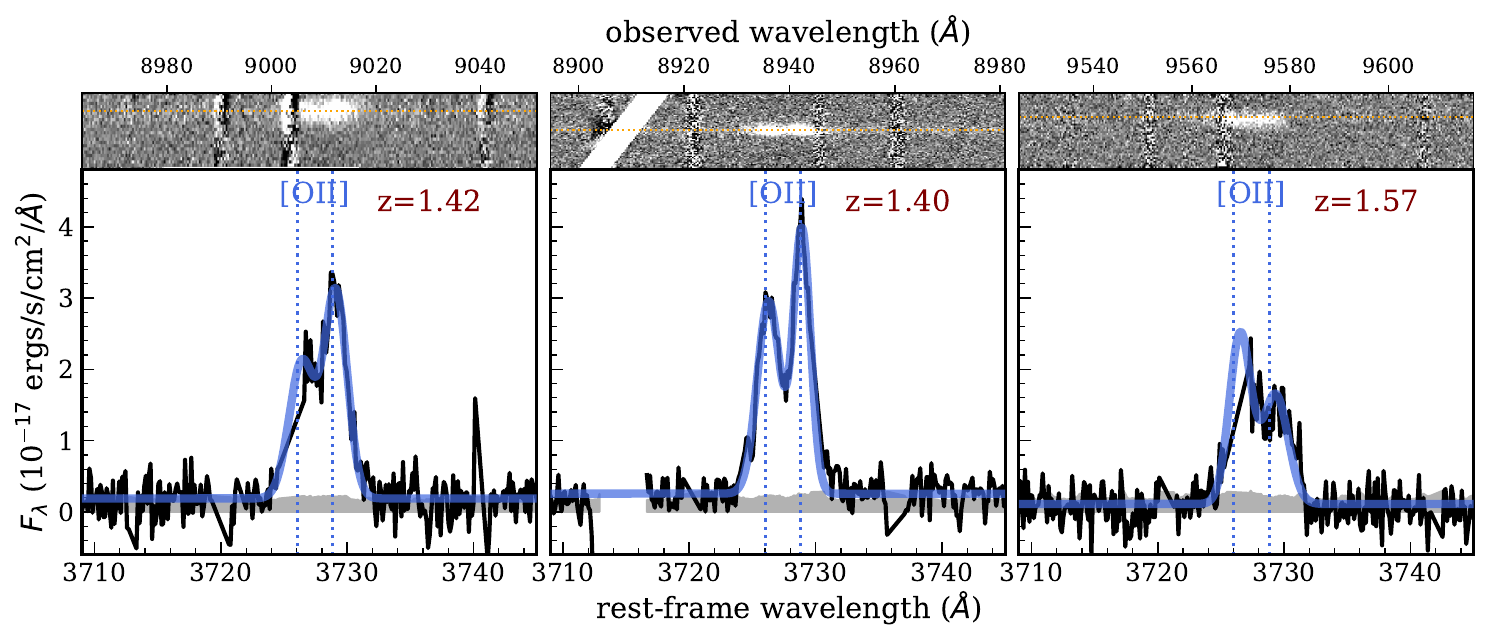}
\caption{Example DEIMOS 2D (top) and 1D (bottom) science spectra of three star-forming galaxies, focusing on the [OII]$\lambda\lambda3726, 3729$ emission-line doublet. Strong sky lines are masked out in the 1D spectra. The best fit is overlaid in blue, with the gray-shaded regions displaying the error spectrum.
\label{fig:Deimos spectra}}
\end{figure*}

To achieve the final flux calibration, the spectra for all objects on a given mask were scaled using a normalization factor derived by comparing the slit star spectrum to their corresponding 3D-\textit{HST} photometric spectral energy distribution (SED), effectively compensating for the slit losses of a point source. The spectra were then corrected for flux losses outside of the slit aperture following the procedure described in \citet{2015ApJ...806..259R} and \citet{2015ApJS..218...15K}. We used the F160W imaging and segmentation map from 3D-\textit{HST} \citep{2014ApJS..214...24S} to identify the pixels associated with the target galaxy and mask out those belonging to nearby sources. We then replaced the masked-out pixels by Gaussian random noise, determined from the average and standard deviation of the background pixel values. Next, we used a Gaussian kernel with FWHM = $\sqrt{\text{FWHM}_{\text{seeing}}^2 - \text{FWHM}_{\text{F160W}}^2}$ to smooth the postage stamp, where $\text{FWHM}_{\text{seeing}}$ represents the seeing, determined from the spatial profile of the slit star measured from the 2D spectrum, and $\text{FWHM}_{\text{F160W}}$ refers to the FWHM of the F160W point spread function. We fitted the smoothed image with a 2D elliptical Gaussian, and calculated the fraction of total source light falling within the slit by integrating the fitted elliptical Gaussian over the slit boundaries. Finally, we scaled each galaxy spectrum by the ratio of the fraction of in-slit light for the slit star relative to that of the galaxy. The scaling factor accounts for the excess slit loss of extended galaxy targets relative to a point source. This approach ensures that the slitloss corrections are performed in a consistent manner to what was done for the near-infrared MOSFIRE spectra from MOSDEF \citep{2015ApJ...806..259R}.
Examples of detected [OII] doublets from the DEIMOS spectroscopy are shown in Figure~\ref{fig:Deimos spectra}.

\subsection{Galaxy properties and emission-line measurements}\label{subsec:Galaxy_properties}
The method for estimating stellar masses is described in  \citet{2015ApJS..218...15K} and \citet{2021ApJ...914...19S}. Briefly, broadband photometry was drawn from the 3D-\textit{HST} survey catalogs \citep{2012ApJS..200...13B, 2014ApJS..214...24S}. Nebular emission line contributions were first subtracted from the photometry in filters covering the rest-frame optical \citep{2021ApJ...914...19S}. Stellar masses were then determined using the SED fitting code \texttt{FAST} \citep{2009ApJ...700..221K}, employing flexible stellar population synthesis models \citep{2009ApJ...699..486C}. The fitting procedure adopted a \citet{2000ApJ...533..682C} dust attenuation curve, solar stellar metallicities, the \citet{2003PASP..115..763C} initial mass function (IMF), and assumed a `delayed-$\tau$' star-formation history (SFR$(t) \propto t \times e^{-t/\tau}$). Although a Small Magellanic Cloud (SMC) extinction law \citep{2003ApJ...594..279G} and subsolar metallicity ($Z_{*}=0.0031$) are suggested to be more appropriate for high-redshift galaxies \citep{2015ApJ...806..259R, 2018ApJ...853...56R, 2020ApJ...899..117S}, galaxies at $z\sim1.5$ typically exhibit metallicities of  12 + log(O/H)~$> 8.5$. In this higher-metallicity regime, the average attenuation curve tends to have a shallower slope \citep{2020ApJ...899..117S}, more closely resembling the \citet{2000ApJ...533..682C} starburst law than the steeper SMC-like curve, thereby justifying our choice.

The fluxes for all emission lines, except [OII]$\lambda\lambda3726,3729$ when covered by DEIMOS, were directly obtained from the MOSDEF survey catalogs\footnote{\url{https://mosdef.astro.berkeley.edu/for-scientists/data-releases/}}\citep{2015ApJ...806..259R, 2015ApJS..218...15K}. For sources for which [OII] was instead covered by DEIMOS, [OII]$\lambda\lambda3726,3729$ fluxes were determined by simultaneously fitting a linear continuum and a double Gaussian profile (see Figure~\ref{fig:Deimos spectra}), with uncertainties estimated by perturbing the science spectrum according to the error spectrum 1000 times and refitting the line in each iteration. The flux uncertainties were then defined as the 68$^{\text{th}}$-percentile width of the resulting distribution.

\begin{figure*}
\includegraphics[width=\textwidth]{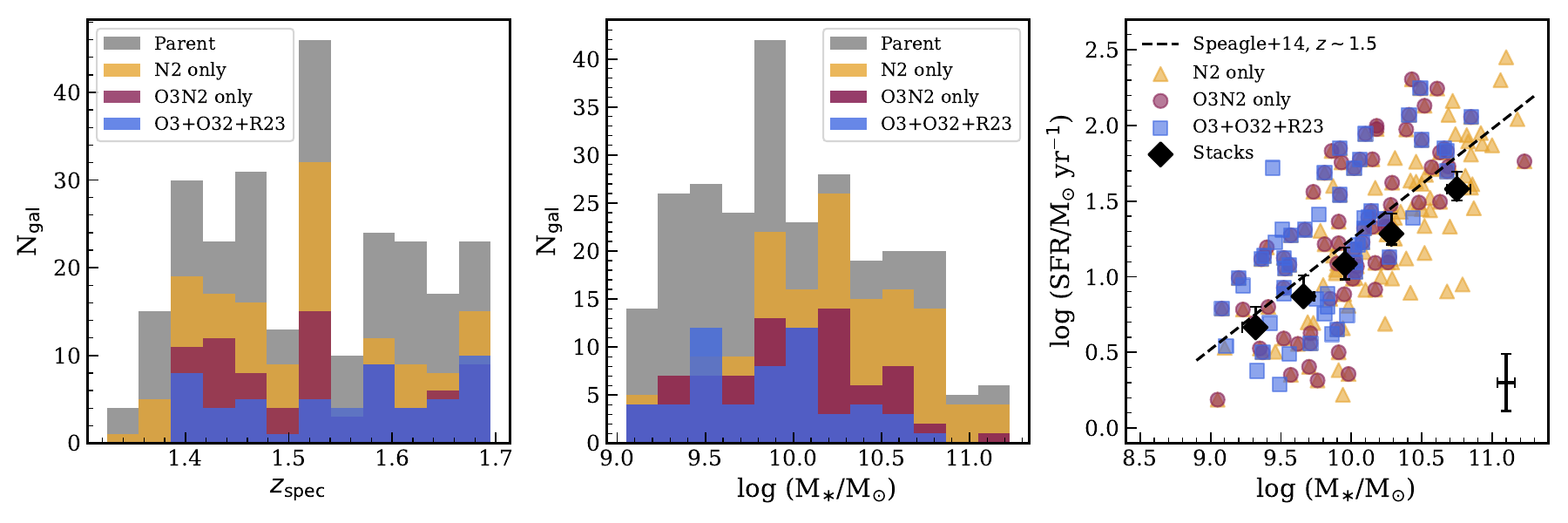}
\caption{\textit{Left}: Redshift distribution of galaxies in our $z\sim1.5$ sample. The colored histograms display the subsamples of galaxies with detections of all lines required for the stated line ratios, described in Table~\ref{tab:stacks_summary}. \textit{Middle}: Stellar mass distribution. \textit{Right}: SFR-$\text{M}_{\ast}$ distribution of the $z\sim1.5$ samples. The dashed line represents the star-forming main sequence from \citet{2014ApJS..214...15S}. The colored triangles, circles, and squares represent the individual galaxies. The values from inferred stacked spectra in stellar mass bins are represented by black diamonds. The error bar in the lower right corner display the median uncertainties of all the individual galaxies.
\label{fig:SFMS}}
\end{figure*}

Emission line fluxes were corrected for dust attenuation assuming the \citet{1989ApJ...345..245C} extinction curve. When both H$\alpha$ and H$\beta$ were detected at S/N$\ge$3, $E(B-V)_{\text{gas}}$ was derived from the H$\alpha$/H$\beta$ ratio assuming an intrinsic ratio of 2.86. If H$\beta$ had S/N$<$3, $E(B-V)_{\text{gas}}$ was estimated from the best-fit stellar population parameters using equation~1 in \citet{2021ApJ...914...19S}.
In the $z\sim1.5$ stacking sample, upon which the main results of this analysis are based, 73/112 galaxies have S/N$\ge$3 for both H$\alpha$ and H$\beta$, while the remaining 39 objects have S/N$<3$ for H$\beta$.
All emission-line ratios were calculated using dust-corrected fluxes. SFR was derived from the dust-corrected H$\alpha$ luminosity using the calibration defined by \citet{2011ApJ...741..124H} converted to a \citet{2003PASP..115..763C} IMF. While galaxies selected to have individual detections of certain lines (e.g., requiring [OII] detections at high M$_{\ast}$; see Section~\ref{Sample selection} and Table~\ref{tab:stacks_summary}) tend to lie above the main sequence in some mass ranges, the stacking sample -- which requires S/N$\ge$3 for H$\alpha$ only -- remains representative of the main-sequence star-forming galaxy population at $z \sim 1.5$, as shown in the right panel of Figure~\ref{fig:SFMS}.

\subsection{Sample selection}\label{Sample selection}
We selected 359 galaxies within the redshift range  $1.30 \leq z \leq 1.70$\footnote{While MOSDEF targeted galaxies with $z_{\text{spec}}$ or $z_{\text{phot}}$ in the range $1.37 \leq z \leq 1.70$, a small number of galaxies with measured $z_{\text{spec}}$ between 1.30 and 1.37 were included in the sample due to slight discrepancies between the measured $z_{\text{spec}}$ and estimated $z_{\text{phot}}$.} from the MOSDEF survey. We rejected galaxies identified as AGN based on X-ray or rest-frame infrared properties, or having [NII]$\lambda$6585/H$\alpha>0.5$ \citep{2015ApJ...801...35C, 2017ApJ...835...27A, 2019ApJ...886...11L}. We also excluded galaxies with stellar masses below $\log(\text{M}_{\ast}/\text{M}_{\odot}) = 9.0$ where MOSDEF is significantly incomplete. These cuts resulted in a total sample of $z\sim1.5$ MOSDEF star-forming galaxies numbering 295 with a median stellar mass of $\log(\text{M}_{\ast}/\text{M}_{\odot}) = 9.91$ and $z_{\text{median}} = 1.53$. We used this sample of star-forming galaxies to define several sub-samples with different emission-line detection requirements: one for producing composite spectra in bins of stellar mass, and others consisting of individual galaxies based on lines necessary for metallicity estimates from different line ratios (see Section~\ref{sec:Z_calculations} for metallicity estimation). We summarize the sub-samples in Table \ref{tab:stacks_summary}. We first required H$\alpha$ S/N$\geq$3, resulting in a sample of 259 galaxies referred to as the parent sample. The additional requirement for wavelength coverage of [OII], H$\beta$, [OIII], and [NII] reduced the sample to a total of 112 galaxies. The median stellar mass and SFR of these 112 galaxies is comparable to that of the full MOSDEF $z \sim 1.5$ sample. We also selected samples of individual galaxies with $\geq 3\sigma$ detections of sets of lines required for different metallicity indicators: 55 galaxies with detections of [OII], H$\beta$, [OIII], and H$\alpha$ to derive the O3, O32, and R23 ratios; 147 galaxies with detections of H$\alpha$ and [NII] to derive N2; and 81 galaxies with detections of H$\beta$, [OIII], H$\alpha$, and [NII] to derive O3N2.
All of these sub-samples have median stellar masses within $\approx$0.2~dex of the parent MOSDEF star-forming galaxy sample and median SFRs consistent with the star-forming main sequence at this redshift (see Table~\ref{tab:stacks_summary}). The redshift and stellar mass distributions of the described galaxy samples are shown in the left and middle panels of Figure~\ref{fig:SFMS}, while the relation between SFR and $\text{M}_{\ast}$ is presented in the right panel.

\begin{deluxetable*}{lccccc}
\tabletypesize{\scriptsize}
\tablewidth{0pt}
\tablecaption{Characteristics of the $z\sim1.5$ star-forming galaxy samples.\label{tab:stacks_summary}}
\tablehead{\colhead{Sample type} & 
\colhead{Wavelength Coverage} & \colhead{Detected Lines ($\geq 3\sigma$)} & \colhead{$\mathrm{N}_{\text{gal}}$} & \colhead{$\log(\text{M}_{\ast}/\text{M}_{\odot})_{\text{med}}$} &
\colhead{$\log(\text{SFR}/\text{M}_{\odot}\text{yr}^{-1})_{\text{med}}$} 
}
\colnumbers
\startdata
Parent Sample & All MOSDEF ($1.3 \leq z \leq 1.7$) & H$\alpha$ & 259 & 9.92 & 1.09 \\
Stacking sample & [OII], H$\beta$, [OIII], H$\alpha$, [NII] & H$\alpha$ & 112  & 9.99 & 1.12  \\
O3+O32+R23 sample & [OII], H$\beta$, [OIII], H$\alpha$ & [OII], H$\beta$, [OIII], H$\alpha$ & 55 & 9.83 & 1.38   \\
N2 sample & H$\alpha$, [NII] & H$\alpha$, [NII] & 147 & 10.17 & 1.45  \\
O3N2 sample & H$\beta$, [OIII], H$\alpha$, [NII] & H$\beta$, [OIII], H$\alpha$, [NII] & 81 & 10.02 & 1.50   \\
\enddata
\end{deluxetable*}

\begin{deluxetable*}{lccccccccccc}
\tabletypesize{\scriptsize}
\tablewidth{0pt}
\tablecaption{Properties of the mass-binned composite spectra for the $z \sim 1.5$ stacking sample.\label{tab:Properties_Stack}}
\tablehead{
\colhead{$\log \left(\frac{\text{M}_{\ast}}{\text{M}_{\odot}}\right)_{\text{med}}$} & \colhead{$\mathrm{N}_{\text{gal}}$} & \colhead{$\log\left(\frac{\text{SFR}}{\text{M}_{\odot}/\text{yr}}\right)_{\text{med}}$} & \colhead{$\log (\text{O3})$} &
\colhead{$\log (\text{O2})$} & \colhead{$\log (\text{O32})$} & \colhead{$\log (\text{R23})$} & \colhead{$\log (\text{N2})$} & \colhead{$\log (\text{O3}\text{N2})$} & \colhead{$\log \left(\frac{\text{O}}{\text{H}}\right)^{\text{O-based}}$} &
\colhead{$\log \left(\frac{\text{O}}{\text{H}}\right)^{\text{O3N2}}$} &
\colhead{$\log \left(\frac{\text{O}}{\text{H}}\right)^{\text{N2}}$}
}
\colnumbers
\startdata
9.32$^{+0.05}_{-0.09}$ & 22 & 0.67$^{+0.13}_{-0.04}$ & 0.56$^{+0.06}_{-0.06}$ & 0.41$^{+0.12}_{-0.02}$ & 0.14$^{+0.02}_{-0.11}$ & 0.86$^{+0.07}_{-0.04}$ & -1.10$^{+0.12}_{-0.06}$ & 1.66$^{+0.09}_{-0.15}$ & 8.37$^{+0.03}_{-0.03}$ &
8.38$^{+0.05}_{-0.05}$ & 8.40$^{+0.05}_{-0.05}$\\
9.66$^{+0.08}_{-0.03}$ & 22 & 0.87$^{+0.14}_{-0.05}$ & 0.41$^{+0.09}_{-0.03}$ & 0.45$^{+0.06}_{-0.09}$ & -0.05$^{+0.13}_{-0.04}$ & 0.79$^{+0.06}_{-0.04}$ & -0.96$^{+0.06}_{-0.09}$ & 1.37$^{+0.16}_{-0.07}$ & 8.50$^{+0.03}_{-0.03}$ & 8.50$^{+0.05}_{-0.05}$ & 8.47$^{+0.03}_{-0.04}$ \\
9.96$^{+0.06}_{-0.03}$ & 22 & 1.09$^{+0.11}_{-0.10}$ & 0.35$^{+0.03}_{-0.09}$ & 0.55$^{+0.06}_{-0.08}$ & -0.21$^{+0.06}_{-0.10}$ & 0.82$^{+0.04}_{-0.07}$ & -0.78$^{+0.07}_{-0.05}$ & 1.12$^{+0.05}_{-0.12}$ & 8.54$^{+0.03}_{-0.03}$ & 8.59$^{+0.03}_{-0.03}$ & 8.56$^{+0.03}_{-0.03}$ \\
10.29$^{+0.04}_{-0.05}$ & 22 & 1.28$^{+0.13}_{-0.07}$ & 0.09$^{+0.08}_{-0.07}$ & 0.49$^{+0.08}_{-0.10}$ & -0.40$^{+0.12}_{-0.08}$ & 0.68$^{+0.07}_{-0.07}$ & -0.61$^{+0.04}_{-0.06}$ & 0.71$^{+0.12}_{-0.10}$ & 8.69$^{+0.03}_{-0.03}$ & 8.76$^{+0.04}_{-0.04}$ & 8.64$^{+0.02}_{-0.02}$ \\
10.75$^{+0.10}_{-0.07}$ & 24 & 1.58$^{+0.12}_{-0.07}$ & 0.09$^{+0.11}_{-0.17}$ & 0.49$^{+0.31}_{-0.13}$ & -0.40$^{+0.09}_{-0.29}$ & 0.66$^{+0.25}_{-0.09}$ & -0.50$^{+0.03}_{-0.04}$ & 0.59$^{+0.13}_{-0.17}$ & 8.70$^{+0.06}_{-0.07}$ & 8.80$^{+0.06}_{-0.05}$ & 8.70$^{+0.02}_{-0.02}$ \\
\enddata
\end{deluxetable*}

\subsection{Composite spectra}\label{subsec:stacking}
\begin{figure*}
\includegraphics[width=\textwidth]{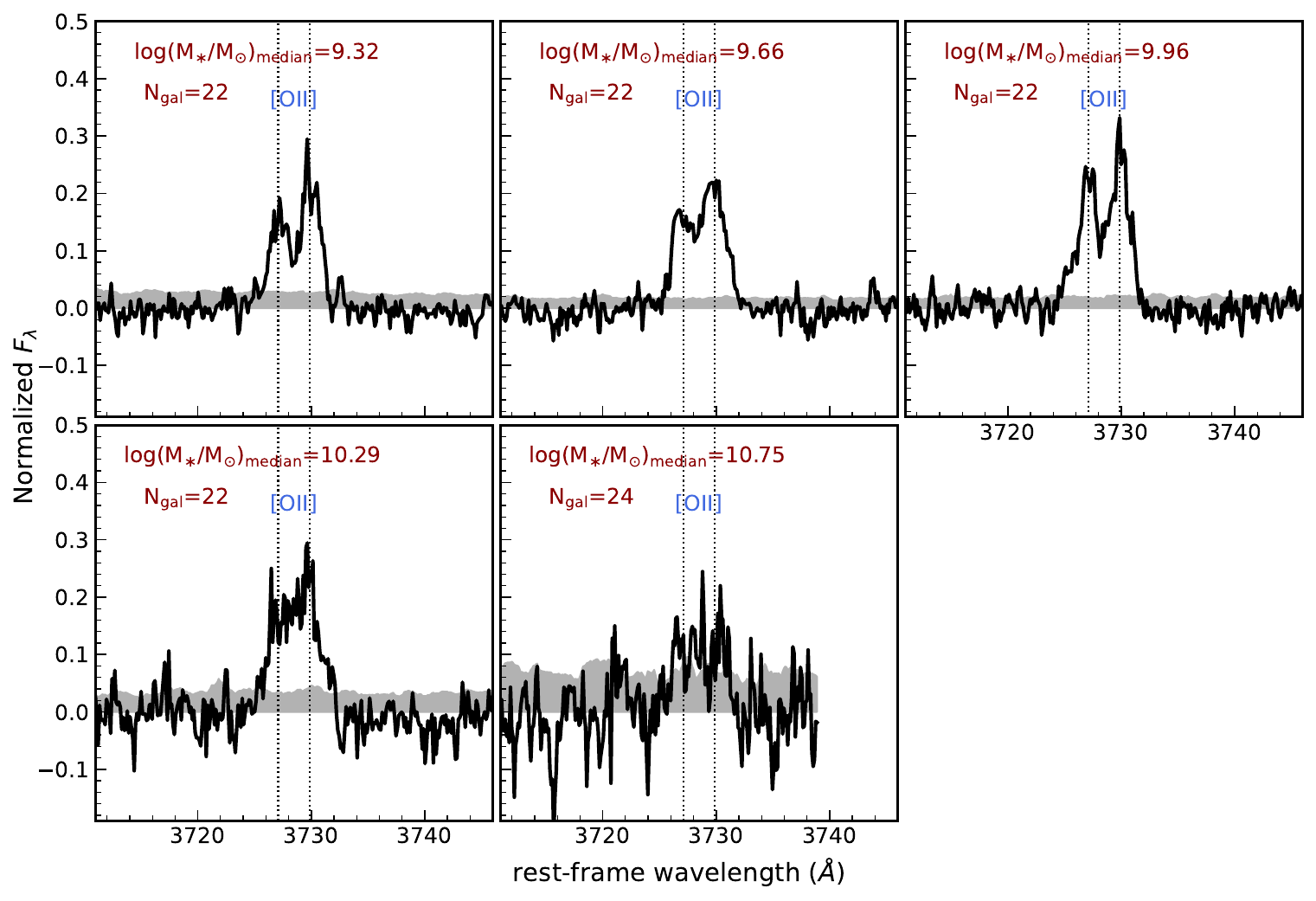}
\caption{Stacked composite spectra in bins of stellar mass with flux normalized by H${\alpha}$, focused on the wavelength range close to [O II]$\lambda\lambda3726, 3729$. The gray-shaded region displays the composite error spectrum.
\label{fig:Stacked spectra}}
\end{figure*}

Constructing composite spectra enables us to include information from galaxies for which weak lines of interest (e.g., H$\beta$, [NII]) were not individually detected. We adopted the stacking method described in \citet{2021ApJ...914...19S} to generate composite spectra in bins of stellar mass for the stacking sample of 112 galaxies with H$\alpha$ S/N$\ge$3 and spectral coverage of [OII], H$\beta$, [OIII], and [NII]. The stacking sample was divided into 5 mass bins, the first four of which at $\log(\text{M}_{\ast}/\text{M}_{\odot}) < 10.5$ have an equal number per bin (22 galaxies), while the final high-mass bin (24 galaxies) at $\log(\text{M}_{\ast}/\text{M}_{\odot}) > 10.5$ covers a region where MOSDEF may not be complete (see the discussion in \citealt{2015ApJS..218...15K, 2021ApJ...914...19S}). Briefly, individual spectra were dust corrected, shifted to rest-frame luminosity density, normalized by H$\alpha$ luminosity, resampled onto a uniform wavelength grid, and median-combined. Stacked line luminosities were measured from Gaussian profile fitting, and uncertainties on line luminosities and line ratios were estimated using a bootstrap Monte Carlo approach, following \citet{2021ApJ...914...19S}. Figure~\ref{fig:Stacked spectra} shows the five resulting composite spectra, focusing on the [OII]$\lambda\lambda3726,3729$ emission lines.

It of note that the [OII] coverage could come from either DEIMOS or MOSFIRE observations, depending on the redshift of the target. Since the DEIMOS spectra have approximately twice the spectral resolution of the MOSFIRE spectra, DEIMOS spectra were smoothed with a Gaussian kernel and resampled during stacking to match the resolution and sampling of MOSFIRE $Y$ band spectra. To ensure that subsets with [OII] coverage from DEIMOS and MOSFIRE could be combined without introducing bias, we compared composite line ratios as a function of M$_{\ast}$ measured from stacked subsamples with [OII] coverage exclusively from either DEIMOS or MOSFIRE. As shown in Appendix~\ref{app:stacking}, we found consistent results from stacks of the two non-overlapping subsets, confirming that sources with [OII] coverage drawn from DEIMOS and MOSFIRE can be combined during stacking.

\subsection{Literature spectroscopic samples}

We utilized published galaxy survey datasets to probe the low-redshift regime at $z < 1$, and higher redshifts at $z=2-4$, complementing the $z\sim1.5$ sample described above. Specifically, we searched the literature for samples of representative star-forming galaxies with published composite spectra or binned sample averages with coverage of at least the [OII], H$\beta$, and [OIII] lines, also utilizing H$\alpha$ and [NII] coverage when available. Stellar masses and SFRs for all samples drawn from the literature were renormalized to a \citet{2003PASP..115..763C} IMF.
Each literature sample was also limited to masses at log M$_{\ast}$/M$_{\odot} > 9.0$ for consistency with the mass range covered by the high-redshift ($z>1$) samples.
The final combination of 6 redshift bins at $z=0.08$, 0.3, 0.8, 1.5, 2.3, and 3.3 probe respective lookback times of 1~Gyr, 3.4~Gyr, 6.8~Gyr, 9.2~Gyr, 10.6~Gyr, and 11.6~Gyr, providing a time sampling of $1-3$~Gyr over the past 85\% of cosmic history.
The properties of the mass-binned literature samples are reported in Appendix~\ref{app:literature} and Table~\ref{tab:Properties_all_z}.

\subsubsection{SDSS -- $z\sim0.08$}\label{SDSS_stacks}
We utilized composite spectra at $z\sim0.08$ for $\sim~200,000$ galaxies from Sloan Digital Sky Survey \citep[SDSS;][]{2000AJ....120.1579Y}, as constructed by \citet{2013ApJ...765..140A}. These stacks are binned in 0.1~dex bins in M$_{\ast}$ and 0.25~dex bins in SFR. To obtain a mass-binned average that avoids potential biases from the inclusion of extreme starburst galaxies, we first select bins lying within $\pm 0.75$\,dex of the star-forming main sequence at $z\sim0$. We exclude any bins containing fewer than 100 galaxies. For the remaining stacks, we compute sample averaged line ratios for each stellar mass interval by averaging the dust-corrected line ratios of the SFR bins at that M$_{\ast}$ weighted by the number of galaxies in each bin. The composite SFR at each M$_{\ast}$ was computed using the same weighted average method. \citet{2011ApJ...730..137Z} and \citet{2024ApJ...964...59L} noted that the MPA-JHU stellar masses for SDSS \citep{2003MNRAS.341...33K,2007ApJS..173..267S} are biased by +0.2~dex relative to the stellar masses used for LEGA-C, DEEP2, and MOSDEF, due to differences in the SED fitting techniques employed. Accordingly, we shift the SDSS stellar masses down by 0.2~dex to ensure consistency with the masses estimated for the higher-redshift samples.

\subsubsection{SHELS -- $z\sim0.3$}\label{SHELS}
We use composite spectra of 2131 star-forming galaxies at $z=0.2-0.38$ from the Smithsonian Hectospec Lensing Survey (SHELS) presented in \citet{2014ApJS..213...35G}. These authors derived stellar masses using the Le Phare software package \citep{2006A&A...457..841I} applied to SDSS five-band photometry, and constructed stacked spectra of star-forming galaxies in 10 bins of M$_{\ast}$. We adopt the median stellar mass in each bin, and measure line fluxes of [OII], H$\beta$, [OIII], H$\alpha$, and [NII] from the published composite spectra by fitting Gaussian profiles to the emission lines, including a 2-component fit for H$\alpha$ and H$\beta$ to account for stellar absorption.
SFRs were not reported for this sample, and could not be measured directly from the stacked spectra since their flux density is normalized. To recover physical flux units, we first fit the relation between $R$-band magnitude and stellar mass for individual SHELS star-forming galaxies. Using the $R$-band filter transmission curve from the Deep Lens Survey, we generated mock photometry by passing the stacked spectra through this response function. We then scaled each stack to match the expected photometric $R$-band flux at the median stellar mass of the bin, thereby placing the stacks onto an absolute flux-calibrated scale. After applying dust correction, we derived SFRs from the dust-corrected H$\alpha$ luminosity for each stack.

\subsubsection{DEEP2 and LEGA-C surveys -- $z\sim0.8$}\label{DEEP2}
We use the line ratios, stellar masses, and SFRs presented in \citet{2011ApJ...730..137Z} for composite spectra in bins of M$_{\ast}$ of $\sim$1350 star-forming galaxies at $z\sim0.8$ from the Deep Extragalactic Evolutionary Probe 2 survey \citep[DEEP2;][]{2013ApJS..208....5N}. This DEEP2 sample spans a mass range of log M$_{\ast}$/M$_{\odot} \sim 9.0-10.5$.

To supplement the DEEP2 sample at higher stellar masses, we use the dust-corrected line ratios, M$_{\ast}$, and SFR measurements for 145 star-forming galaxies at $z\sim0.8$ spanning log M$_{\ast}$/M$_{\odot} \sim 10.0-11.0$ presented in \citet{2024ApJ...964...59L}, drawn from the Large Early Galaxy Astrophysics Census \citep[LEGA-C;][]{2021ApJS..256...44V} survey.
The sample is then binned in stellar mass, and average values of SFR and emission line ratios are computed within each bin. Following \citet{2024ApJ...964...59L}, based on their similarity in mean redshift and in line ratios in the region of overlapping stellar mass, we combine the DEEP2 and LEGA-C survey composites into a single $z\sim0.8$ sample for the remainder of the analysis.
Both DEEP2 and LEGA-C spectra cover [OII], H$\beta$, and [OIII] at this redshift, but do not cover H$\alpha$ and [NII], which are redshifted into the NIR.
 
\subsubsection{MOSDEF survey -- $z\sim2.3$ and $z\sim3.3$}

We adopt published stellar masses, SFRs, and dust-corrected line ratios for composite spectra of 280 star-forming galaxies at $z=2.1-2.6$ and 115 star-forming galaxies at $z=3.0-3.8$ from the MOSDEF survey \citep{2021ApJ...914...19S}.
These samples span log~M$_{\ast}$/M$_{\odot} \sim9.0-11.0$, but are potentially incomplete at masses above log M$_{\ast}$/M$_{\odot} =10.5$ \citep{2015ApJS..218...15K,2021ApJ...914...19S}.

\subsection{Combined $z=0-4$ sample properties}

\begin{figure}
\includegraphics[width=\columnwidth]{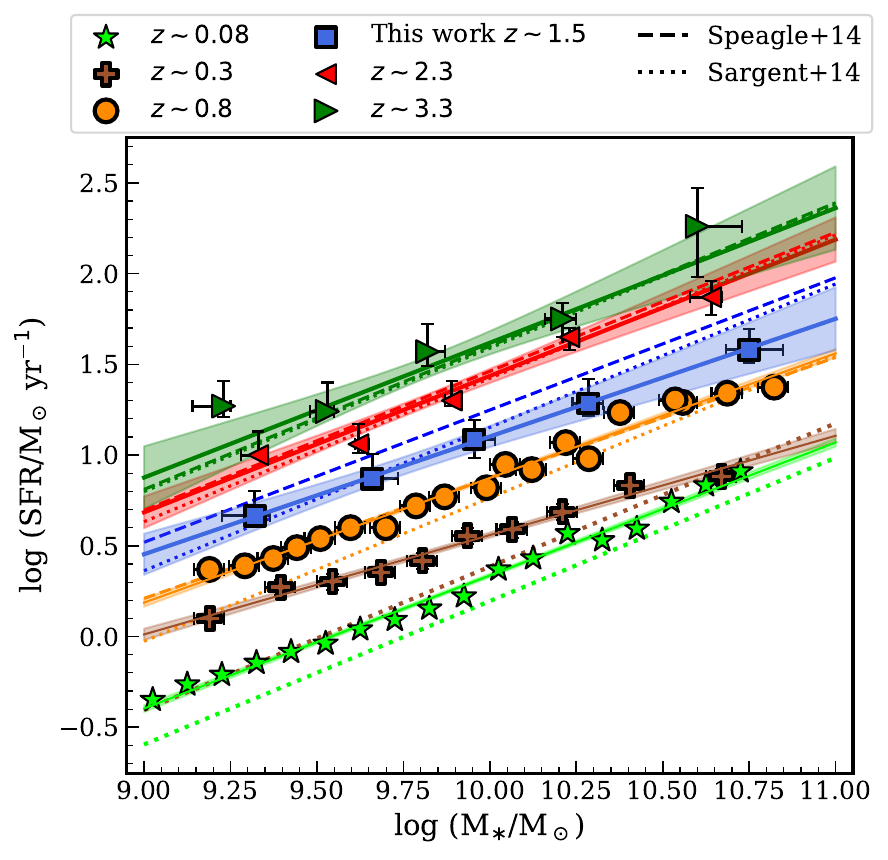}
\caption{SFR-$\text{M}_{\ast}$ relations defined by the mass-binned samples across all 6 redshift bins included in this study.
Solid colored lines indicate the best-fit relations at each redshift with shaded regions representing the $1\sigma$ uncertainties on the fits, showing similar slopes but an increase in SFR at fixed mass with increasing redshift.
Dotted lines show the star-forming main sequence parameterization of \citet{2014ApJ...793...19S}, evaluated at the median redshift of each sample.
Dashed lines represent the main sequence of \citet{2014ApJS..214...15S}, evaluated at the median redshifts of the $z=0.8-3.3$ samples.
\label{fig:SFMS_all_z}}
\end{figure}

Figure~\ref{fig:SFMS_all_z} presents the SFR-$\text{M}_{\ast}$ distribution of the composites across all 6 redshift bins included in this study. Best-fit relations to these data are overlaid as solid lines of the same colors, with shaded regions represent the $1\sigma$ uncertainty.
The dotted lines represent the star-forming main sequence parametrization from the \citet{2014ApJ...793...19S} evaluated at the mean redshift of each sample. We find agreement with the \citet{2014ApJ...793...19S} relation within 0.2~dex in SFR at fixed M$_\ast$ for the vast majority of the composites across the full redshift range.
The exceptions are the lowest-mass $z\sim3.3$ bin, as well as the $z\sim0.3$ bins below $10^{10}$~M$_\odot$, which fall higher than the \citet{2014ApJ...793...19S} relation.
This offset may indicate that the $z\sim0.3$ sample is biased below $10^{10}$~M$_\odot$, though this scenario seems unlikely since SHELS maintains high completeness below this mass \citep{2014ApJS..213...35G}.
Another possibility is that our approach to calibrate the $z\sim0.3$ stacks to an absolute flux scale (Sec.~\ref{SHELS}) has overestimated the correction at low mass.
It is thus possible that we have overestimated the SFRs of the lowest-mass $z\sim0.3$ bins by up to 0.3~dex.

We also compare the $z=0.8-3.3$ samples to the main sequence parameterization of \citet{2014ApJS..214...15S} (dashed lines).\footnote{\citet{2014ApJS..214...15S} excluded samples at $z\lesssim0.3$ from their analysis. Thus, their fit to the redshift-dependent star-forming main sequence may not be reliable if applied at the redshifts of our $z\sim0.08$ and $z\sim0.3$ samples. See also the discussion in \citet{2018A&A...615A...7L}.}
We find that our four highest-redshift samples agree with the \citet{2014ApJS..214...15S} relation within $\sim0.2$~dex in SFR at fixed M$_\ast$, except for the lowest-mass $z\sim3.3$ bin.
We conclude that all 6 of our mass-binned samples are representative of the typical main-sequence star-formation galaxy population at each redshift across the majority of the mass range log M$_{\ast}$/M$_{\odot} = 9.0-11.0$.

\section{METALLICITY DERIVATIONS} \label{sec:Z_calculations}
We estimated metallicities using dust-corrected ratios of the rest-optical lines [OII], H$\beta$, [OIII], H$\alpha$, and [NII]. This set of emission lines provides a number of line ratios from which gas-phase metallicities can be inferred using strong-line calibrations.
The lines [OII], H$\beta$, and [OIII] are covered in all 6 redshift bins in this study, such that we can use ratios of these lines uniformly across the redshift range $z=0-4$. We estimate gas-phase oxygen abundance following a method similar to the one described in \citet{2021ApJ...914...19S}.
We use the combination of the O3, O32, and R23 line ratios to infer our fiducial set of metallicities, which we refer to as ``O-based.''
These line ratios have the advantage of only involving lines of O and H to determine O/H, avoiding the dependence of [NII]-based indicators on N/O. To account for evolving ISM ionization conditions, we use a different set of strong-line calibrations for low- and high-redshift samples. For the $z < 1$ galaxy samples, we adopted the metallicity calibrations from \citet{2017MNRAS.465.1384C} and \citet{2020MNRAS.491..944C}, based on direct-method oxygen abundances measured from stacked SDSS spectra. For $z > 1$ galaxy samples, we used the high-redshift analog calibrations from \citet{2018ApJ...859..175B}, which are based on direct-method metallicities of extreme local galaxies that reflect the ISM physical conditions of $z \sim 2$ galaxies. Specifically, these analogs share similar emission line properties with high redshift galaxies, particularly occupying similar positions in the [NII]-BPT diagram \citep[e.g.,][]{2014ApJ...795..165S,2016ApJ...816...23S}. It is worth noting that the \citet{2018ApJ...859..175B} relations are calibrated within the metallicity range of $7.8 < 12+\log\text{(O/H)} < 8.4$. Consequently, application at $12+\log\text{(O/H)} > 8.4$ requires extrapolation. Although this introduces some uncertainty, \citet{2021ApJ...914...19S} found this method to be a better match for high-$z$ galaxies compared to applying $z \sim 0$ calibrations to $z > 1$ samples. In particular, applying the \citet{2020MNRAS.491..944C} calibration to high-$z$ galaxies typically results in oxygen abundances that are, on average, underestimated by $\sim0.1$~dex compared to the \citet{2018ApJ...859..175B} analog calibration.

We calculated metallicities via a $\chi^2$ minimization using:
\begin{equation}
    \chi^2 = \sum\limits_i\frac{(R_{\text{obs}, i} - R_{\text{cal}, i}(x))^2}{(\sigma_{\text{obs}, i}^2 + \sigma_{\text{cal}, i}^2)}
\end{equation}
where the summation over $i$ represents the set of line ratios (O3, O32, and R23 for O-based metallicities), $R_{\text{obs}, i}$ is the logarithm of the $i^{\text{th}}$ observed line ratio, $R_{\text{cal}, i}$ is the logarithmic $i^{\text{th}}$ line ratio of the calibration at $x = 12 + \log(\text{O/H})$, $\sigma_{\text{obs}, i}$ is the uncertainty in the $i^{\text{th}}$ observed line ratio, and $\sigma_{\text{cal}, i}$ is the intrinsic scatter in the $i^{\text{th}}$ line ratio at fixed O/H of the calibration.
The best-fit metallicity corresponds to the value that minimizes the expression above. When deriving metallicities for mass-binned composites, $\sigma_{\text{cal}, i}$ is divided by $\sqrt{N}$ where $N$ is the number of galaxies in the bin. We adopted $\sigma_{\text{cal}, i}$ for the \citet{2018ApJ...859..175B} calibrations from \citet{2021ApJ...914...19S}.
We determined the uncertainties in the metallicity estimates by employing a Monte Carlo approach, perturbing the observed line ratios by randomly adding Gaussian noise based on their uncertainties and repeating the fit 300 times to generate a distribution of O/H values. The final 1$\sigma$ uncertainties were taken to be half of the 16$^{\text{th}}$-to-84$^{\text{th}}$ percentile width of this distribution. Table~\ref{tab:Properties_Stack} summarizes the obtained galaxy properties, emission line ratio fluxes and inferred metallicities. 

The N2 and O3N2 line ratios are also available for 4 of the redshift bins, at $z\sim 0.08$, 0.3, 1.5, and 2.3. While we view these [NII]-based line ratios as potentially less reliable due to their additional sensitivity to N/O, they are nonetheless very commonly employed metallicity indicators at both low and high redshifts. Accordingly, we also infer metallicities from either N2 or O3N2 for comparison. As above, we use the \citet{2020MNRAS.491..944C} calibrations at $z<1$. At $z>1$, we use the high-redshift analog N2 and O3N2 calibrations of  \citet{2018ApJ...859..175B} as renormalized by \citet{2023ApJ...942...24S} to better match the O-based metallicities at $z\sim2$.

The formal uncertainties on metallicities in many of the low-redshift bins are very small ($\ll0.01$~dex) due to the larger number of galaxies in these samples, especially for the SDSS composites. As a consequence of inverse-variance weighting, the low-redshift samples would dominate over those at higher redshift when fitting a redshift-dependent functional form to the mass-metallicity relation. Furthermore, systematic uncertainties on composite line ratios and metallicities are far larger than such small formal errors. As a result, we set a metallicity error floor of 0.01~dex for all samples throughout the analysis,\footnote{Our results do not significantly change if we vary this metallicity error floor between 0.01 and 0.04~dex.} adopting this value for the error on 12+log(O/H) if the formal uncertainty was smaller. 

\section{Results} \label{sec:Results}

\subsection{Emission line ratios versus stellar mass}
\begin{figure*}
\includegraphics[width=\textwidth]{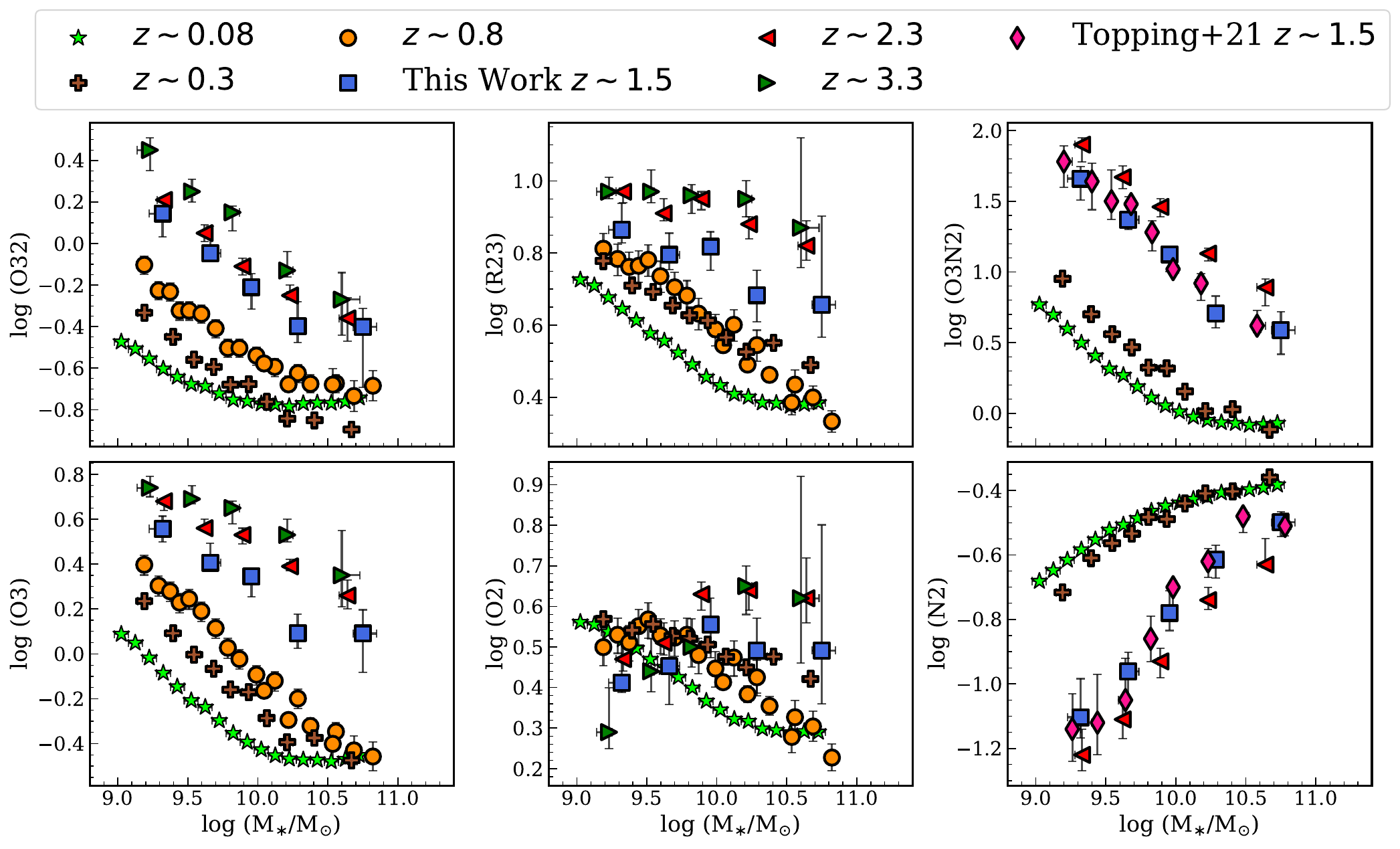}
\caption{Rest-optical emission-line ratios as a function of stellar mass of mass-binned composites for the star-forming galaxy samples at 6 redshift intervals included in this analysis.
All line ratios have been corrected for dust reddening.
\label{fig:Lines_ratios}}
\end{figure*}

We first explore the empirical evolution of rest-optical emission-line ratios at fixed stellar mass over $z=0-4$. Figure~\ref{fig:Lines_ratios} shows O32, O3, R23, and O2 as a function of M$_{\ast}$ for all 6 redshift bins, as well as O3N2 and N2 for the 4 bins in which [NII] is covered. At fixed redshift, we find that O32, O3, R23, and O3N2 decrease with increasing M$_{\ast}$, while N2 increases. These trends suggest that metallicity increases and ionization parameter decreases with increasing M$_{\ast}$ at each redshift.
O2 transitions from an anticorrelation with M$_{\ast}$ at $z<1$ to a positive correlation at $z>2$. O2 is double-valued as a function of O/H, with a turnover metallicity of 12+log(O/H)~$\approx 8.5$, suggesting that the majority of the $z<1$ sample has metallicities higher than this value while the $z>2$ samples lie largely at lower metallicities.
The low-mass flattening in O3 and R23 at $z>2$ can similarly be ascribed to the double-valued nature of these line ratios with O/H, with respective turnover points of 12+log(O/H$)\approx 8.0$ and 8.2.
At $z<1$, the line ratios flatten at high mass, reflecting the well-known high-mass asymptotic behavior of the mass-metallicity relation.
For O3N2 and N2, we also compare to the $z\sim1.5$ stacked spectra from \citet{2021MNRAS.506.1237T} based on a larger MOSDEF sample of $\sim$250 galaxies, which shows good agreement with the trends in our smaller $z\sim1.5$ sample due to the requirement of [OII] coverage.

At fixed stellar mass, O32, O3, R23, and O3N2 display a nearly ubiquitous increase with redshift, while N2 decreases with increasing redshift at fixed stellar mass.
These empirical trends suggest that metallicity decreases and the degree of nebular ionization increases with increasing redshift at fixed M$_{\ast}$, as found by earlier studies with fewer redshift bins \citep[e.g.,][]{2017ApJ...835...88K,2021ApJ...914...19S}.
O2 in contrast increases with increasing redshift at log(M$_{\ast}$/M$_\odot)>10.0$ while decreasing with increasing redshift below that mass, which is, again, due to the double-valued nature of this ratio as a function of O/H.
The rate of line ratio evolution is slower at the high-mass end, with  O32, O3, R23, O3N2, and N2 displaying little evolution at log(M$_{\ast}$/M$_\odot)>10.5$ until $z>1$, suggesting metallicity evolution at fixed M$_{\ast}$ is faster at low masses and slower at high masses. 
The highest-mass bins 
(log(M$_{\ast}$/M$_\odot)>10.2$) in the $z\sim0.3$ dataset display slight deviations from these trends for O32 and R23 that may have been introduced due to inaccurate stellar absorption correction or potential composite AGN contamination at such high masses. Specifically, at high masses, the O32 values at $z\sim0.3$ are lower than expected -- falling below the $z\sim0$ sequence -- when they are predicted to lie between the $z\sim0$ and $z\sim0.8$ datasets. Likewise, the high-mass $z\sim0.3$ R23 values are higher than those observed at $z\sim0.8$, again deviating from the expected redshift progression. 
Overall, all of these rest-optical line ratios qualitatively suggest that O/H is positively correlated with M$_{\ast}$ at each redshift and decreases with increasing redshift at fixed M$_{\ast}$, and that a high-mass flattening of O/H is present at $z<1$, asymptoting to approximately the same metallicity at these redshifts.

\subsection{The mass-metallicity relation}\label{sec:MZR}
\begin{figure*}
\includegraphics[width=\textwidth]{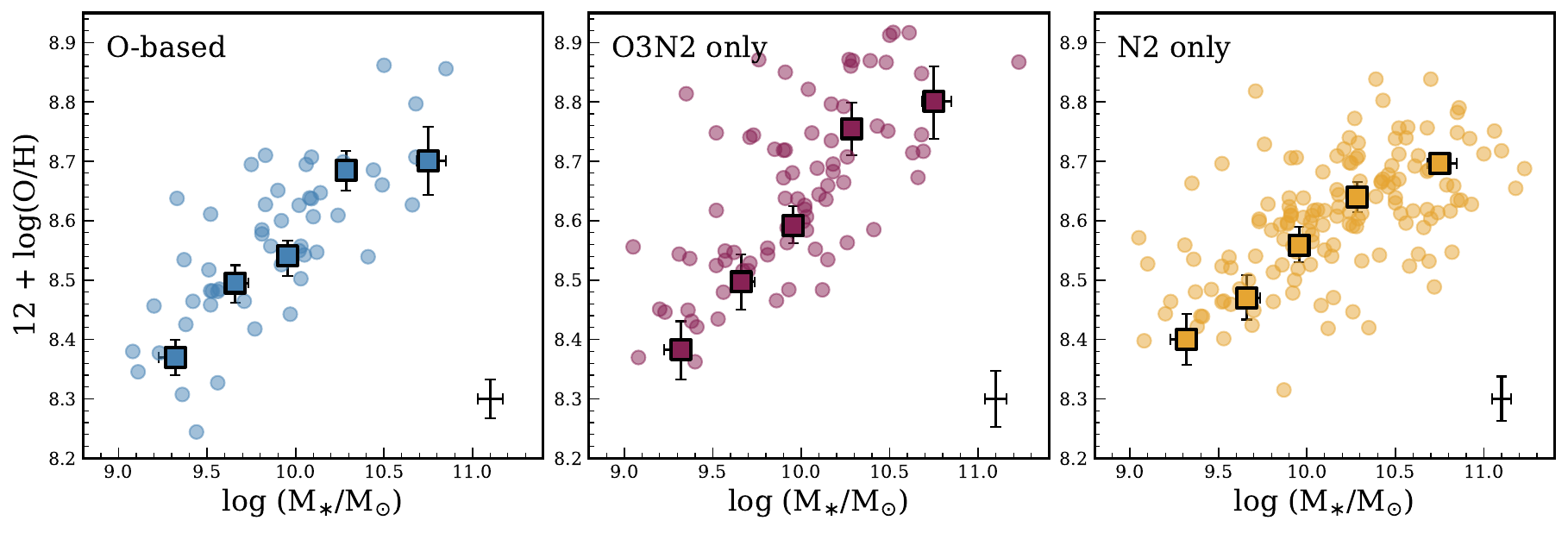}
\caption{The mass-metallicity relation for $z \sim 1.5$ star-forming galaxies based on O-based (left panel), O3N2 (middle panel) and N2 (right panel) line ratio diagnostics, converted to metallicities using the analog calibrations from \citet{2018ApJ...859..175B}. The circles and large squares represent individual galaxies and stacked spectra, respectively. The error bars in the lower right corner of each panel display the median uncertainties of the individual galaxies.
\label{fig:MZR}}
\end{figure*}

In Figure \ref{fig:MZR}, we show the mass-metallicity relation for individual galaxies and stacked spectra for our sample at $z\sim1.5$ incorporating new [OII] measurements, with metallicities derived using O-based ratios (left); O3N2 (middle); and N2 (right) as described in Sec.~\ref{sec:Z_calculations}.
The Spearman correlation coefficients for the set of individual galaxies across the three cases are 0.74, 0.69, and 0.57, with p-values $\ll 10^{-10}$, demonstrating a highly significant correlation between $\text{M}_{\ast}$ and O/H at $z\sim1.5$, regardless of which line ratio is used to determine O/H.
Metallicities derived from the mass-binned composite spectra reflect the average trends observed in the individual $z\sim1.5$ galaxies.
For all three cases, the lower four mass bins display a trend that can be represented as a power-law increase of O/H with $\text{M}_{\ast}$.
The highest-mass bin shows tentative evidence of flattening, though this behavior is inconclusive due to the uncertainties.

\begin{figure*}
\includegraphics[width=\textwidth]{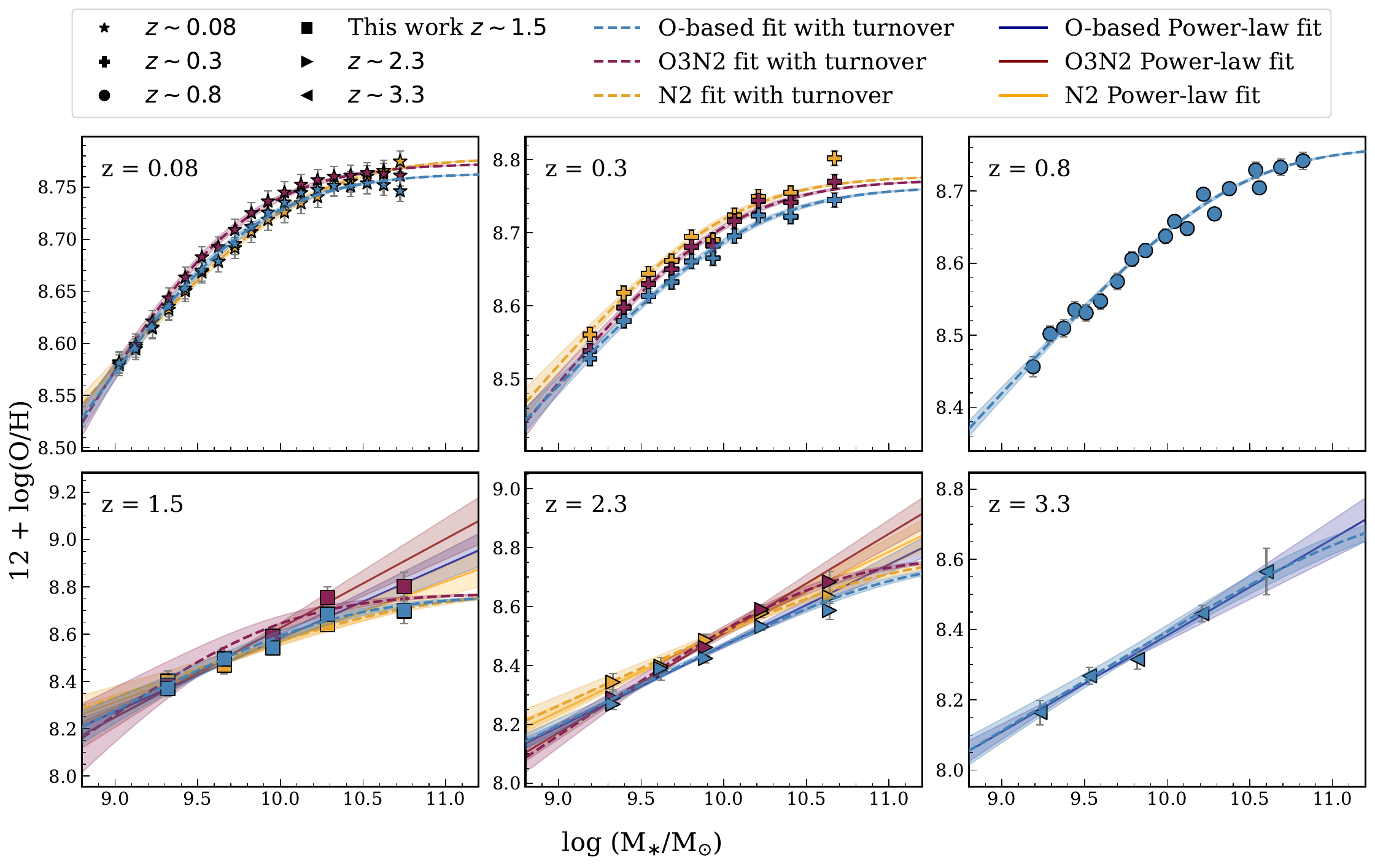}
\caption{Mass-metallicity relations for the 6 redshift bins based on mass-binned composites, color-coded by the line ratio(s) used to derive metallicity.
Blue, magenta, and yellow represent metallicities based on the O-based, O3N2, and N2 line ratio diagnostics, respectively.
Dashed lines display the best-fit MZR functional form that includes high-mass flattening (eq.~\ref{eq:2020_Curti_MZR_z}).
Solid lines represent the best-fit simple power-law for the $z=1.5-3.3$ samples (eq.~\ref{eq:Power_law}).
For both cases, the shaded regions denote the $1\sigma$ uncertainty bounds on the fits.
Best-fit parameters are reported in Tabs.~\ref{tab:BF-MZR_O3O2}, \ref{tab:BF-MZR_O3N2}, and \ref{tab:BF-MZR_N2}.
\label{fig:MZR_all_3}}
\end{figure*}

Figure~\ref{fig:MZR_all_3} compares the metallicity estimates derived using the three strong-line diagnostics (O-based, represented by blue; O3N2, represented by magenta; and N2, represented by yellow) for the composite spectra in all 6 redshift bins, excluding O3N2 and N2 at $z\sim0.8$ and $z\sim3.3$ where [NII] measurements are not available. This comparison enables a direct assessment of how the choice of metallicity indicator affects the shape of the derived MZR. Specifically, this figure illustrates how the slope and normalization of the MZR differ depending on which line ratios are used to calculate the metallicities. We find that O/H estimates from all three line diagnostics are generally consistent with one another, displaying offsets of $<0.05$~dex in most cases. However, systematic variation is present at high masses ($\geq10^{10.3}\text{M}_{\odot}$), with estimates from O3N2 and N2 resulting in higher asymptotic O/H than for the O-based case.

To quantify differences in slope and normalization, we fit the MZR for each redshift and line ratio combination using equation~2 in \citet{2020MNRAS.491..944C} given by: 
\begin{equation}\label{eq:2020_Curti_MZR_z}
     \text{MZR}(\text{M}_{\ast}) = Z_{0} - \gamma/\beta \times \log(1 + [\text{M}_{\ast}/\text{M}_{0}]^{-\beta}),
\end{equation}
where $Z_{0}$ is the high-mass asymptotic metallicity, $\gamma$ is the low-mass power-law slope, and $\beta$ controls the sharpness of the turnover that occurs at mass M$_0$. The fits were performed using an inverse-variance weighted non-linear least squares minimization via the \texttt{curve\_fit} routine in \texttt{SciPy}.
Above $z\sim0.3$, $\beta$ and M$_0$ are not well constrained by the data, which do not reach high enough masses to resolve a full flattening. Instead, we fix $\beta=1.2$, the $z\sim0$ value found by \citet{2020MNRAS.491..944C} that also fits our local SDSS sample well. For each strong-line diagnostic, $Z_{0}$ at each redshift is fixed to the value obtained by fitting the MZR at $z=0.08$, corresponding to $Z_{0}$ of 8.76, 8.77, and 8.78 for O-based, O3N2, and N2 metallicities, respectively. The low-mass slope $\gamma$ and turnover mass M$_0$ thus remain free parameters for each fit. The best-fit coefficients are given in Tables~\ref{tab:BF-MZR_O3O2}, \ref{tab:BF-MZR_O3N2} and \ref{tab:BF-MZR_N2}. The dashed lines and shaded regions of corresponding colors in Figure~\ref{fig:MZR_all_3} represent the best-fit relations and their $1\sigma$ uncertainty bounds.  

\begin{deluxetable}{lcc}[h]
\tabletypesize{\scriptsize}
\tablecaption{Best-fit MZRs to Composite Spectra with O-based metallicities from the O3, O32, and R23 line ratios. \label{tab:BF-MZR_O3O2}}
\tablehead{}
\startdata
\multicolumn{3}{c}{\textbf{Using Power-law eq.~\ref{eq:Power_law}}} \\
\hline
\colhead{Sample} & \colhead{$\gamma$} & \colhead{$Z_{10}$} \\
\hline
$z \sim 1.5$ & $0.31 \pm 0.04$ & $8.58  \pm 0.02$ \\
$z \sim 2.3$ & $0.28 \pm 0.02$ & $8.47 \pm 0.01$ \\
$z \sim 3.3$ & $0.27 \pm 0.04$ & $8.38 \pm 0.02$ \\
\hline
\hline
\multicolumn{3}{c}{\textbf{Using eq.~\ref{eq:2020_Curti_MZR_z} with fixed $\beta=1.2$ and $Z_{0}=8.76$}} \\
\hline
\colhead{Sample} & \colhead{$\gamma$} & \colhead{$\log\left(\frac{\rm M_{0}}{\rm M_\odot}\right)$} \\
\hline
$z \sim 0.08$ & $0.24 \pm 0.03$ & $9.75 \pm 0.09$ \\
$z \sim 0.3$  & $0.24 \pm 0.03$ & $10.13 \pm 0.08$ \\
$z \sim 0.8$  & $0.25 \pm 0.01$ & $10.37 \pm 0.04$ \\
$z \sim 1.5$  & $0.36 \pm 0.09$ & $10.39 \pm 0.22$ \\
$z \sim 2.3$  & $0.28 \pm 0.02$ & $11.05 \pm 0.11$ \\
$z \sim 3.3$  & $0.28 \pm 0.04$ & $11.32 \pm 0.22$ \\
\hline
\hline
\multicolumn{3}{c}{\textbf{Redshift evolution using eq.~\ref{eq:2023_MZR_z} with fixed $\beta=1.2$}} \\
\hline
\colhead{Sample} & \colhead{$Z_{0}$} & \colhead{$\gamma$} \\
\hline
All & $8.76 \pm 0.01$ & $0.28 \pm 0.01$ \\
\hline
 & \colhead{$m_0$} & \colhead{$m_1$} \\
\hline
 & $9.55 \pm 0.05$ & $2.75 \pm 0.08$ \\
\hline
\enddata
\end{deluxetable}

\begin{deluxetable}{lcc}[h]
\tabletypesize{\scriptsize}
\tablecaption{Best-fit MZRs to Composite Spectra with metallicities based on the O3N2 line ratio. \label{tab:BF-MZR_O3N2}}
\tablehead{}
\startdata
\multicolumn{3}{c}{\textbf{Using Power-law eq.~\ref{eq:Power_law}}} \\
\hline
\colhead{Sample} & \colhead{$\gamma$} & \colhead{$Z_{10}$} \\
\hline
$z \sim 1.5$ & $0.38 \pm 0.06$ & $8.63  \pm 0.02$ \\
$z \sim 2.3$ & $0.34 \pm 0.03$ & $8.51 \pm 0.01$ \\
\hline
\hline
\multicolumn{3}{c}{\textbf{Using eq.~\ref{eq:2020_Curti_MZR_z} with fixed $\beta=1.2$ and $Z_{0}=8.77$}} \\
\hline
\colhead{Sample} & \colhead{$\gamma$} & \colhead{$\log\left(\frac{\rm M_{0}}{\rm M_\odot}\right)$} \\
\hline
$z \sim 0.08$ & $0.28 \pm 0.04$ & $9.65 \pm 0.09$ \\
$z \sim 0.3$  & $0.29 \pm 0.04$ & $9.95 \pm 0.09$ \\
$z \sim 1.5$  & $0.52 \pm 0.20$ & $10.06 \pm 0.27$ \\
$z \sim 2.3$  & $0.38 \pm 0.05$ & $10.62 \pm 0.13$ \\
\hline
\hline
\multicolumn{3}{c}{\textbf{Redshift evolution using eq.~\ref{eq:2023_MZR_z} with fixed $\beta=1.2$}} \\
\hline
\colhead{Sample} & \colhead{$Z_{0}$} & \colhead{$\gamma$} \\
\hline
All & $8.77 \pm 0.01$ & $0.40 \pm 0.04$ \\
\hline
 & \colhead{$m_0$} & \colhead{$m_1$} \\
\hline
 & $9.40 \pm 0.08$ & $2.19 \pm 0.11$  \\
\hline
\enddata
\end{deluxetable}

\begin{deluxetable}{lcc}[h]
\tabletypesize{\scriptsize}
\tablecaption{Best-fit MZRs to Composite Spectra with metallicities based on the N2 line ratio. \label{tab:BF-MZR_N2}}
\tablehead{}
\startdata
\multicolumn{3}{c}{\textbf{Using Power-law eq.~\ref{eq:Power_law}}} \\
\hline
\colhead{Sample} & \colhead{$\gamma$} & \colhead{$Z_{10}$} \\
\hline
$z \sim 1.5$ & $0.26 \pm 0.05$ & $8.57  \pm 0.02$ \\
$z \sim 2.3$ & $0.27 \pm 0.03$ & $8.51 \pm 0.01$ \\
\hline
\hline
\multicolumn{3}{c}{\textbf{Using eq.~\ref{eq:2020_Curti_MZR_z} with fixed $\beta=1.2$ and $Z_{0}=8.78$}} \\
\hline
\colhead{Sample} & \colhead{$\gamma$} & \colhead{$\log\left(\frac{\rm M_{0}}{\rm M_\odot}\right)$} \\
\hline
$z \sim 0.08$ & $0.19 \pm 0.02$ & $10.07 \pm 0.08$ \\
$z \sim 0.3$  & $0.27 \pm 0.04$ & $9.95 \pm 0.09$ \\
$z \sim 1.5$  & $0.25 \pm 0.05$ & $10.83 \pm 0.19$ \\
$z \sim 2.3$  & $0.26 \pm 0.03$ & $11.03 \pm 0.16$ \\
\hline
\hline
\multicolumn{3}{c}{\textbf{Redshift evolution using eq.~\ref{eq:2023_MZR_z} with fixed $\beta=1.2$}} \\
\hline
\colhead{Sample} & \colhead{$Z_{0}$} & \colhead{$\gamma$} \\
\hline
All & $8.78 \pm 0.01$ & $0.25 \pm 0.01$ \\
\hline
 & \colhead{$m_0$} & \colhead{$m_1$} \\
\hline
 & $9.72 \pm 0.06$ & $2.63 \pm 0.08$  \\
\hline
\enddata
\end{deluxetable}

Since the highest-mass bin in each sample at $z\ge1.5$ may be significantly incomplete and potentially biased, we also fit the high-redshift samples ($z=1.5-3.3$) using a power law form excluding the highest-mass bin, following the method used by \citet{2021ApJ...914...19S}:
\begin{equation}\label{eq:Power_law}
    12 + \log(\text{O/H}) = \gamma \times m_{10} + Z_{10},
\end{equation}
where $\gamma$ is the slope, $m_{10} = \log(\text{M}_{\ast}/10^{10}\text{M}_{\odot})$ and $Z_{10}$ is the metallicity at $10^{10}\ \text{M}_{\odot}$. The best-fit power-law parameters are given in Tables~\ref{tab:BF-MZR_O3O2}, \ref{tab:BF-MZR_O3N2} and \ref{tab:BF-MZR_N2}. The solid lines and corresponding shaded regions in Figure~\ref{fig:MZR_all_3} represent the best-fit power-law relations and $1\sigma$ uncertainties for the three $z\ge1.5$ samples. 

Overall, the slopes of the fits for O/H estimates from all three line diagnostics are generally consistent with each other at fixed redshift, despite slight variations in steepness. The fit from O3N2 favors a slightly steeper slope relative to MZRs derived from O-based ratios, while N2 exhibits a shallower slope relative to the others. Nearly all cases are consistent with a low-mass slope of $\gamma\approx0.25-0.35$ regardless of redshift, metallicity indicator used, or functional form (power law vs.\ including high-mass flattening). The best-fit turnover masses are consistent between the O-based and O3N2 cases, but slightly higher when fit to the N2 metallicities.

We thus find that the slope and normalization of the MZR, and its redshift evolution, are not strongly sensitive to the choice of metallicity indicator when using appropriate redshift-dependent strong-line calibrations. It is also of note that the turnover mass, M$_0$, uniformly increases with increasing redshift.
This fact, alongside the similarity in slope measured independently at each redshift, suggests that the evolving MZR could be well-fit with a functional form that adopts a constant slope at all redshifts and a redshift-dependent turnover mass.
This evolution in M$_0$ is effectively equivalent to a redshift-dependent normalization in the power-law framework — that is, an increase in redshift corresponds to a decrease in normalization when the MZR is modeled as a single-slope power law.
A similar conclusion was reached by \citet{2014ApJ...791..130Z}, in which the evolving MZR at $z=0-1.5$ was fit by a functional form with a redshift invariant low-mass slope and a turnover mass that increased with increasing redshift.

\begin{figure*}[htb]
\includegraphics[width=1.\textwidth]{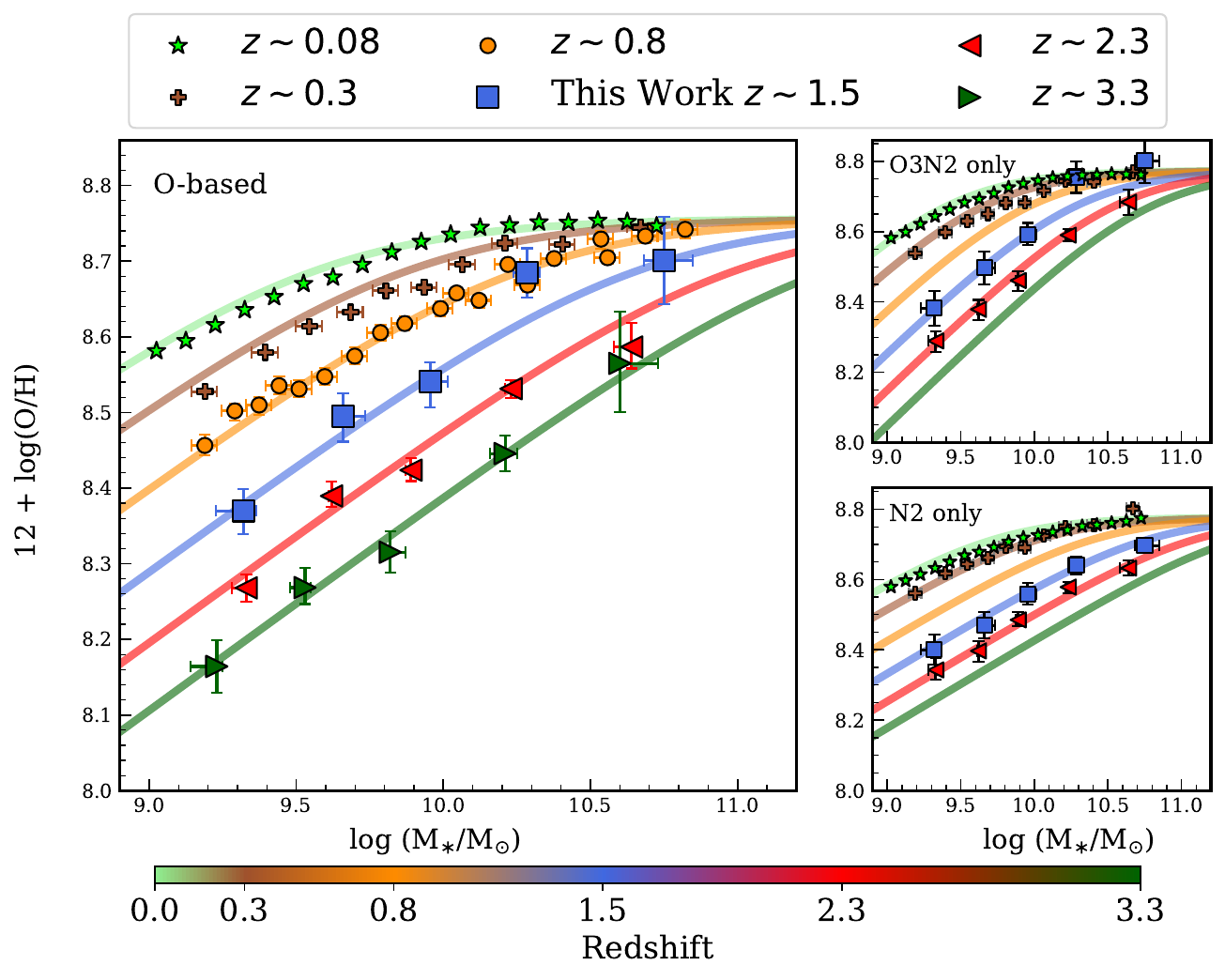}
\caption{The evolution of the mass-metallicity relation over $z\sim0-4$ for mass-binned composites in 6 redshift intervals.
The left panel presents the MZR for metallicities derived from the O-based line ratios (O3, O32, and R23).
The upper-right and lower-left panels dispaly metallicities based on the O3N2 and N2 ratios, respectively, and exclude the $z\sim0.8$ and $z\sim3.3$ bins for which [NII] measurements were not available.
In each panel, the colored lines represent the best-fit redshift-dependent MZR using the functional form of eq.~\ref{eq:2023_MZR_z} and evaluated at the redshift of each sample, with best-fit parameters reported in Tabs.~\ref{tab:BF-MZR_O3O2}, \ref{tab:BF-MZR_O3N2}, and \ref{tab:BF-MZR_N2}.
\label{fig:MZR_all_z}} 
\end{figure*} 

Leveraging the information gained from fitting each redshift bin separately, we now address the redshift evolution of the MZR combining the samples at all 6 redshifts.
The left panel of Figure~\ref{fig:MZR_all_z} presents the evolving MZR based on metallicities derived using the O-based line ratios uniformly across all 6 redshift bins spanning $z\sim0-4$.
The right panels display MZR evolution using metallicities based on the O3N2 (top) and N2 (bottom) ratios for the 4 redshift bins where [NII] was covered. As discussed in Sec.~\ref{sec:Z_calculations}, the left panel is our fiducial case as it uses only lines of O and H to infer O/H, while the right panels have an additional sensitivity to the N abundance. We fit the redshift evolution of the MZR with the parameterization proposed by \citet{2023ApJ...942...24S} which assumes a redshift-invariant low-mass slope and high-mass asymptotic metallicity, but a redshift-dependent turnover mass: 
\begin{equation}\label{eq:2023_MZR_z}
     \text{MZR}(\text{M}_{\ast},z) = Z_{0} - \gamma/\beta \times \log(1 + [\text{M}_{\ast}/\text{M}_{0}(z)]^{-\beta}),
\end{equation}
with parameters as in equation~\ref{eq:2020_Curti_MZR_z}, except that the turnover mass is redshift dependent with the parameterization $\log(\text{M}_{0}(z)/\text{M}_\odot) = m_0 + m_1(1+z)$.
We adopt a fixed $\beta=1.2$, and leave all other parameters free while fitting the combined data from all redshift bins.
The best-fit parameters for the O-based, O3N2, and N2 cases are given in Tables~\ref{tab:BF-MZR_O3O2}, \ref{tab:BF-MZR_O3N2}, and \ref{tab:BF-MZR_N2}, respectively.
Solid lines in Figure~\ref{fig:MZR_all_z} display the best-fit redshift-dependent relation evaluated at the redshift of each sample with matching colors.

It can be seen from Figure~\ref{fig:MZR_all_z} that the assumption of a non-evolving low-mass slope and asymptotic metallicity, in combination with a turnover mass that increases with increasing redshift, provides a good match to the observational constraints for nearly every mass bin at each redshift.
The most significant outliers are the second-highest mass bin at $z\sim1.5$ when using O-based metallicities, and the two highest mass bins for O3N2 at $z\sim1.5$. The latter may indicate that the \citet{2018ApJ...859..175B} analog calibrations that are specifically tuned to match $z\sim2.3$ galaxies provide less self-consistent results at $z\sim1.5$.
The offset of the second-highest O-based $z\sim1.5$ bin may be driven by sample variance, as there are only 22 galaxies in that bin.
It is ultimately desirable to have a strong-line calibration set that is a continuous function of redshift, or of other properties related to ISM ionization conditions, but such an advance awaits future work and data sets.
The samples at $z<1$ show a clear preference for a non-evolving asymptotic metallicity, but the $z>1$ data sets lack the statistics and completeness at high masses to further test this conclusion, requiring wider-area near-IR spectroscopic surveys.

\subsection{The fundamental metallicity relation}
\begin{figure*}
\includegraphics[width=\textwidth]{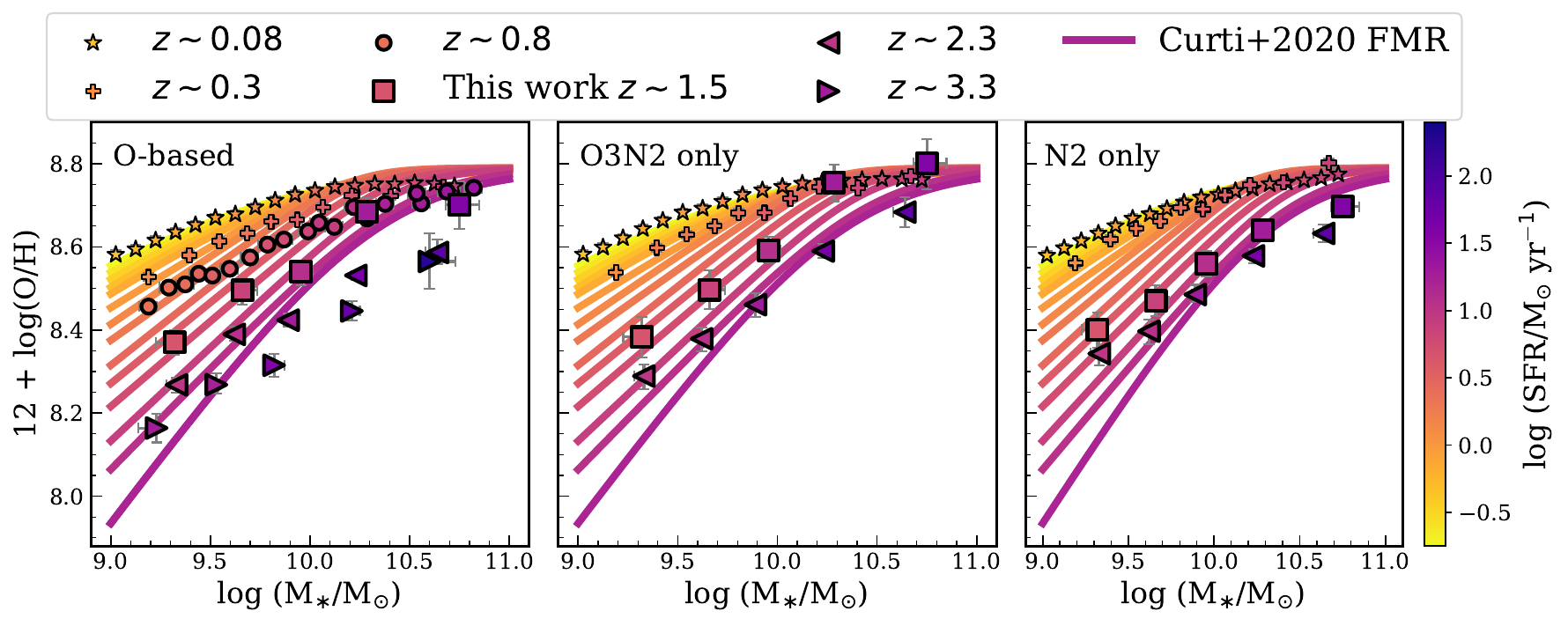}
\caption{Metallicity values estimated from O-based (left panel), O3N2 (middle panel) and N2 (right panel) line ratio diagnostics vs.\ stellar mass, color-coded by SFR. The solid lines illustrate the $z\sim0$ best fit FMR given in Table 5 of \citet{2020MNRAS.491..944C} for comparison, displaying lines of constant SFR following the same color coding.
\label{fig:Curti_comparison}}
\end{figure*}

We now turn to the relation among metallicity (O/H), $\text{M}_{\ast}$, and SFR---the FMR, considered ``fundamental'' due to its proposed invariance with redshift. Figure~\ref{fig:Curti_comparison} plots O/H as a function of $\text{M}_{\ast}$, color-coded by SFR for all of the redshift bins and with O/H derived using the three different metallicity indicators.
The solid colored lines in the background denote the best-fit $z\sim0$ FMR taken from \citet{2020MNRAS.491..944C}, displaying lines of constant SFR color-coded in the same way as the composite spectra in our sample. In this plot, a non-evolving FMR would manifest as matching colors at fixed $\text{M}_{\ast}$ and O/H between the $z\sim0$ lines and composites at different redshifts. We note that the \citet{2020MNRAS.491..944C} SDSS sample extends only up to $\log(\text{SFR}) \sim 1.2$, such that higher SFRs are not represented in these $z\sim0$ curves and comparisons of the local FMR to high-redshift data sets with higher SFRs requires extrapolation. The composites at each redshift generally show good agreement in O/H with the $z \sim 0$ fits at matched $\text{M}_{\ast}$ and SFR, though there are some visible deviations.
The high-mass $z\sim1.5$ bins in O3N2 once again display a significant disagreement. However, no other mass bins at any redshift are obviously offset by such a large amount from the \citet{2020MNRAS.491..944C} FMR.
This good visual agreement suggests little FMR evolution out to $z\sim4$, in agreement with previous studies \citep{2010MNRAS.408.2115M,2010A&A...521L..53L,2019A&A...627A..42C,2020MNRAS.491..944C,2021ApJ...914...19S}.

\begin{figure*}
\includegraphics[width=\textwidth]{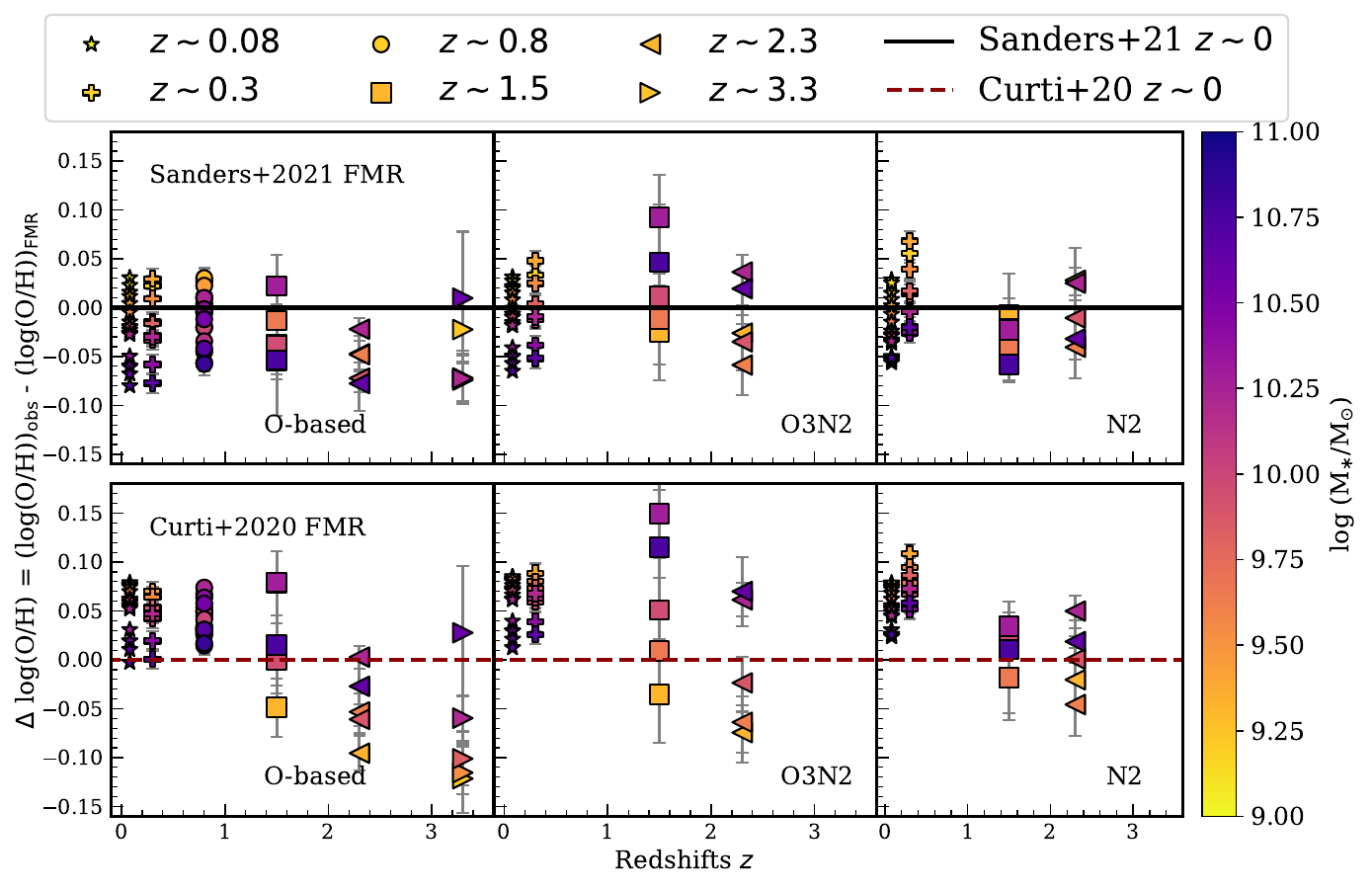}
\caption{O/H residuals relative to the best-fit $z\sim0$ FMR relations from \citet{2021ApJ...914...19S} (black solid line; upper panel) and \citet{2020MNRAS.491..944C} (dashed dark red line; lower panel) as a function of redshift, color-coded by their stellar mass. The metallicities are derived using the O-based (left panel), O3N2 (middle panel), and N2 (right panel) line ratios. 
\label{fig:Residual}}
\end{figure*}

To investigate the question of FMR evolution of the mean star-forming population in further detail, we plot the metallicity residuals from the predictions of a parameterized $z\sim0$ FMR as a function of redshift.
We compare to two parameterizations of the $z\sim0$ FMR in Figure~\ref{fig:Residual}.
In the top row, the FMR parameterization is taken from equation 10 of \citet{2021ApJ...914...19S}.
In the bottom row, we use the FMR defined in equation 5 of \citet{2020MNRAS.491..944C}. The figure is color-coded by M$_{\ast}$. The vertical axis shows the offset in O/H for each composite bin relative to the FMR parameterization evaluated at matched M$_{\ast}$ and SFR. We evaluate residuals separately for each of the three metallicity indicators. The mean of the residuals and root-mean-square (RMS) scatter values relative to the two FMR parameterization are summarized in Table~\ref{tab:Scatter_RMS}.
Our purpose is to evaluate whether commonly-used parameterized FMR forms from the literature can describe gas-phase metallicities over a wide range of redshifts.
To this end, we adopt the parameterizations from these two past studies without alteration despite slight differences in methodologies.
As a result, the $z\sim0.08$ stacks used in this work do not agree exactly with the predictions of these $z\sim0$ FMR functions.
We note that the metallicity derivation techniques in \citet{2021ApJ...914...19S} more closely match the methods used in this work.

\begin{deluxetable}{lccc}[h]
\tabletypesize{\scriptsize}
\tablecaption{Mean and RMS of log(O/H) residuals from the \citet{2021ApJ...914...19S} and \citet{2020MNRAS.491..944C} FMR paramterizations, for different line diagnostics and redshifts (see Figure~\ref{fig:Residual}).\label{tab:Scatter_RMS}}
\tablehead{}
\startdata
\multicolumn{4}{c}{\textbf{\citet{2021ApJ...914...19S} FMR}} \\
\hline
\colhead{Diagnostic} & \colhead{Redshift} & \colhead{Mean residual} & \colhead{RMS Scatter} \\
\hline
O32    & 0.08& $-0.022$ & $0.038$ \\
       & 0.3 & $-0.020$ & $0.038$ \\
       & 0.8 & $-0.010$ & $0.028$ \\
       & 1.5 & $-0.024$ & $0.036$ \\
       & 2.3 & $-0.053$ & $0.057$ \\
       & 3.3 & $-0.046$ & $0.058$ \\
\hline
O3N2   & 0.08& $-0.012$ & $0.032$ \\
       & 0.3 & $-0.001$ & $0.029$ \\
       & 1.5 & $0.023$  & $0.048$ \\
       & 2.3 & $-0.013$ & $0.038$ \\
\hline
N2     & 0.08& $-0.022$ & $0.035$ \\
       & 0.3 & $0.014$  & $0.033$ \\
       & 1.5 & $-0.030$ & $0.034$ \\
       & 2.3 & $-0.006$ & $0.029$ \\
\hline
\hline
\multicolumn{4}{c}{\textbf{\citet{2020MNRAS.491..944C} FMR}} \\
\hline
\colhead{Diagnostic} & \colhead{Redshift} & \colhead{Mean residual} & \colhead{RMS Scatter} \\
\hline
O32    & 0.08& $0.052$  & $0.057$ \\
       & 0.3 & $0.044$ & $0.048$ \\
       & 0.8 & $0.041$ & $0.044$ \\
       & 1.5 & $0.011$  & $0.042$ \\
       & 2.3 & $-0.047$ & $0.057$ \\
       & 3.3 & $-0.074$ & $0.092$ \\
\hline
O3N2   & 0.08& $0.062$  & $0.066$ \\
       & 0.3 & $0.063$ & $0.065$ \\
       & 1.5 & $0.058$  & $0.089$ \\
       & 2.3 & $-0.006$ & $0.061$ \\
\hline
N2     & 0.08& $0.052$  & $0.055$ \\
       & 0.3 & $0.078$ & $0.079$ \\
       & 1.5 & $0.005$  & $0.021$ \\
       & 2.3 & $0.001$  & $0.033$ \\
\hline
\enddata
\end{deluxetable}

The mass-binned composites generally display close agreement with the \citet{2021ApJ...914...19S} FMR, with most stacks and the mean offset at all 6 redshifts lying within 0.05~dex in O/H across all redshifts and regardless of which line ratio is used. The offset from the \citet{2020MNRAS.491..944C} FMR is slightly larger, but still typically $\Delta \log(\text{O}/\text{H}) \lesssim 0.1$~dex.
There are systematic trends between the O/H residuals and M$_{\ast}$ present for stacks at $z\sim0.08$, 0.3, and 0.8, which likely stem from differences in the metallicity calibrations and methodologies used to derive the $z\sim0$ FMR in these past studies relative to our methods.
However, all panels in Figure~\ref{fig:Residual} collectively suggest that there is no strong FMR evolution (i.e., $\Delta\log(\mathrm{O/H})<0.1$) between $z\sim0$ and $z\sim4$, regardless of which FMR parameterization is assumed or which of these three metallicity indicators is used. This analysis confirms that FMR non-evolution holds over the $z=0.3-1.5$ range that was not explored in previous studies.

\section{Discussion} \label{sec:Discussions}

\subsection{Evolution of the MZR}

In the previous section, we uniformly derived the MZR for mass-binned averages of representative star-forming galaxy samples spanning $z=0-3.8$ in 6 redshift bins, probing metallicity evolution over the past 12~Gyr with $1-3$~Gyr time sampling (Figure~\ref{fig:MZR_all_z}).
For the first time, we derived metallicities from the same set of line ratios for all 6 redshift bins (namely, O3, O32, and R23; ``O-based'') using strong-line calibrations that reflect the evolution of typical ISM ionization conditions from low ($z<1$) to high ($z>1$) redshifts.
For comparison, we also inferred metallicities from either O3N2 or N2 for the 4 redshift bins in which [NII] measurements were available.
This homogeneous methodology is designed to minimize the systematics of metallicity comparison across redshift to more robustly characterize metallicity evolution out to $z\sim4$ with finer time sampling than past studies.

We identify three main results from our MZR evolution analysis: (1) the lack of significant MZR slope evolution out to $z\sim4$, when using the same line diagnostics to estimate metallicity at all redshifts; (2) the evolution toward lower O/H at fixed M$_{\ast}$ at a rate that is constant with redshift as $d\log(\text{O/H})/dz \approx-0.1$ at masses below the turnover mass; and (3) the lack of evidence for any redshift evolution of the high-mass asymptotic metallicity of the MZR, when it is parametrized in a form with an asymptotic metallicity.

\begin{figure*}[htb]
\includegraphics[width=1.\textwidth]{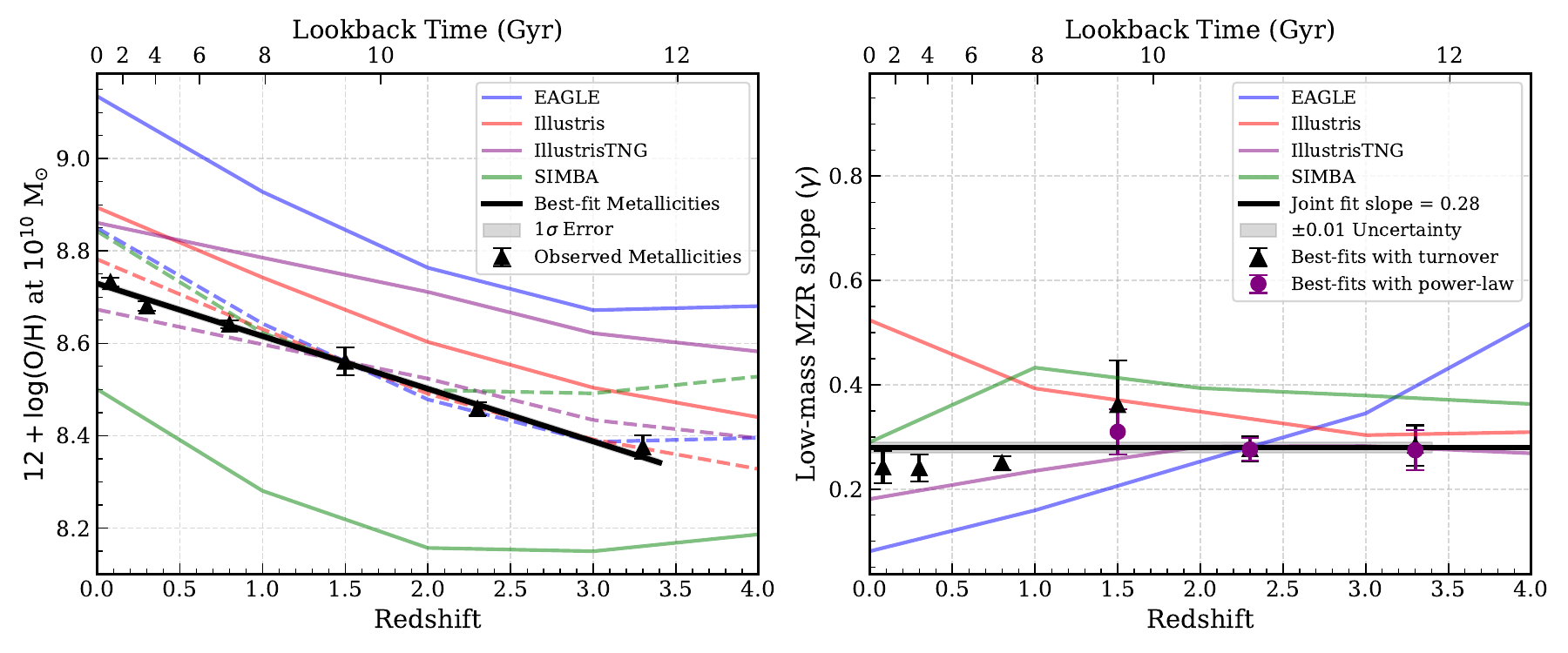}
\caption{\textit{Left:} Redshift evolution of O/H at $10^{10}$~M$_{\odot}$ based on fits to the individual redshift bins using O-based metallicities  (Fig.~\ref{fig:MZR_all_3} and Tab.~\ref{tab:BF-MZR_O3O2}).
The black line shows a best-fit linear function with $d\log(\text{O/H})/dz=-0.11\pm0.01$.
Solid colored lines show predictions from cosmological galaxy formation simulations \citep[EAGLE, Illustris, IllustrisTNG, and SIMBA;][]{2025MNRAS.536..119G}, while dashed colored lines represent the same simulation results renormalized to match the observations at $z=1.5$.
\textit{Right:} Redshift evolution of the low-mass slope of the MZR from O-based metallicities.
Black points show the values from the functional form including high-mass flattening (eq.~\ref{eq:2020_Curti_MZR_z}), while the purple points display results from the power-law fit for the highest three redshift bins (eq.~\ref{eq:Power_law}).
The observational constraints are consistent within $\approx1\sigma$ with a single slope that does not vary out to $z\sim3.3$, found to be $\gamma=0.28\pm0.01$ in the joint fit to all 6 redshift bins (black line; Tab.~\ref{tab:BF-MZR_O3O2}).
Solid colored lines show low-mass MZR slopes measured from the same simulations as in the left panel.
\label{fig:Z10_gamma_evolution}} 
\end{figure*}

Our results provide strong evidence for little to no evolution in the low-mass slope of the MZR over $z\sim0-4$ at masses down to $10^{9} \text{M}_{\odot}$, based on both fits to individual redshift samples (Fig.~\ref{fig:MZR_all_3}) and the joint fit to all redshift bins (Fig.~\ref{fig:MZR_all_z}).
In the right panel of Figure~\ref{fig:Z10_gamma_evolution}, we plot the best-fit low-mass slope ($\gamma$) as a function of redshift from fits to the individual redshift bins using the fiducial O-based metallicities.
It can be seen that the best-fit slopes at all redshifts are $\approx1\sigma$ consistent with each other and with the slope from the joint fit to all redshift bins, $\gamma=0.28\pm0.01$ (black line).
The same result generally holds for the fits based on O3N2 or N2 metallicities with slightly steeper ($\gamma=0.40\pm0.04$) or shallower ($\gamma=0.25\pm0.01$) slopes, respectively.
Regardless of the metallicity indicator adopted, the low-mass MZR appears to maintain a constant mass-dependence of approximately $\text{O/H}\propto \rm M_*^{0.3}$ from $z=0$ up to at least $z\sim3.3$.
This is similar to the slope found in some studies \citep[e.g.][]{2020MNRAS.491..944C,2020A&A...634A.107Y,2021ApJ...914...19S,2021MNRAS.505..903C,2024MNRAS.532.3102S}, but shallower than the steep slopes of $\gamma\approx0.5-0.7$ found by others \citep[e.g.][]{2004ApJ...613..898T,2006ApJ...644..813E,2010MNRAS.408.2115M,2011MNRAS.417.2962P,2013ApJ...765..140A,2014ApJ...792...75Z,2014ApJ...791..130Z,2017ApJ...847...18Z}.
Analyses using modern empirical $T_e$-based calibrations, as we do here, tend to find a slope consistent with our result, while a steeper slope and higher normalization are common when using photoionization-model based calibrations \citep[e.g.][]{2004ApJ...613..898T,2008ApJ...681.1183K,2010MNRAS.408.2115M}.
The steep slope found by \citet{2013ApJ...765..140A}, based on direct-method metallicities measured from stacked SDSS spectra in bins of M$_{\ast}$, may be driven by systematic biases toward lower-than-average metallicities at low stellar masses ($<10^9\ \text{M}_\odot$). These biases likely stem from their emission-line selected sample, which favors extreme, high-sSFR systems, and from their choice to normalize stacks by the rest-optical continuum. This normalization scheme gives greater weight to high equivalent-width objects in the composite spectra--sources that tend to have lower O/H and more extreme ISM conditions.

In analytic gas-regulator or equilibrium chemical evolution models \citep[see e.g.,][]{2008MNRAS.385.2181F, 2011MNRAS.417.2962P, 2012MNRAS.421...98D, 2013ApJ...772..119L,2014ApJ...791..130Z, 2015MNRAS.449.3274F}, the low-mass slope of the MZR is generally set by the inverse of the mass-scaling of either the gas fraction ($f_{\text{gas}}=\rm M_{\text{gas}}/M_*$) or the outflow metal loading factor ($\zeta_\text{out}=\frac{Z_{\text{out}}}{Z_{\text{ISM}}}\frac{\dot{M}_{\text{out}}}{\text{SFR}}$), depending on which factor is larger.
If the metallicity of the outflows ($Z_{\text{out}}$) is equal to the ISM metallicity ($Z_{\text{ISM}}$), then the metal loading factor simplifies to the more common mass loading factor $\eta_{\text{out}}=\frac{\dot{M}_{\text{out}}}{\text{SFR}}$.
The apparently redshift-invariant MZR slope of $\gamma=0.28\pm0.01$ suggests that the same physical process acts to govern this slope over $z\sim0-4$, and that the mass-scaling of this process does not significantly change over that time.
If this process is taken to be metal-loaded outflows, then our constraints imply $\zeta_\text{out}\propto \rm M_*^{-0.3}$.
At $z\sim0$, observations have shown that the gas fractions are low enough that metal removal via outflows dominates in setting the MZR slope \citep[e.g.,][]{2004ApJ...613..898T,2011MNRAS.417.2962P,2013ApJ...765..140A}, in agreement with direct measurements of outflow loading \citep{2017MNRAS.469.4831C,2018MNRAS.481.1690C}.
The non-evolving MZR slope implies outflows maintain status as the dominant mechanism governing the MZR at all redshifts out to at least $z\sim3.3$.
We also note that our best fit slope of $\gamma\approx0.3$ agrees with the expectation from simple momentum-driven wind models \citep[e.g.][]{1986ApJ...303...39D,2005ApJ...618..569M,2011MNRAS.417.2962P}, whereas energy-driven winds predict a slope twice as steep.

As discussed in \citet{2021ApJ...914...19S}, due to the steep rise in $f_{\text{gas}}$ at fixed $\rm M_*$ with increasing redshift \citep[e.g.][]{2020ARA&A..58..157T}, if outflow loading does not also significantly increase with redshift then a point may be reached at $z>4$ where gas fraction begins to dominate instead, potentially leading to a change in the MZR slope at early times.
Observations by {\it JWST} are now probing metallicities in this early-Universe regime.
Early results at $z\sim4-10$ have measured slopes consistent to slightly shallower than what we find here at $z=0-4$:
$\gamma=0.17\pm0.03$ \citep{2024A&A...684A..75C}, $\gamma=0.25\pm0.03$ \citep{2023ApJS..269...33N}, and $\gamma=0.27\pm0.02$ \citep{2025ApJ...978..136S}.
However, these higher-redshift samples still lack both comparable size and a robust evaluation of representativeness relative to the $z<4$ samples employed here, and employ much broader redshift bins.
As the {\it JWST} spectroscopic archive grows, future analyses will be able to extend the uniform approach adopted here to higher redshifts, with this work providing a new benchmark for metallicity evolution from the present day up through Cosmic Noon.

We show the redshift evolution of metallicity at a fixed mass of $10^{10}\ \text{M}_\odot$ in the left panel of Figure~\ref{fig:Z10_gamma_evolution}, using O-based metallicities.
The observational constraints are well-described by a relation that linearly decreases with increasing redshift as $d\log(\text{O/H})/dz =-0.11\pm0.01$ (black line), with every redshift bin within $1\sigma$ of this best-fit line.
We thus find a rate of evolution toward lower metallicity at fixed mass that is invariant with redshift over $z=0-3.3$. While redshift offers a convenient observational axis, it is ultimately lookback time that reflects the physical timescale over which galaxies evolve. Recasting the metallicity evolution in terms of lookback time can provide a more physically meaningful perspective on galaxy chemical enrichment timescales. The linear decrease in log(O/H) as a function of redshift implies that the rate of enrichment is not uniform with time, but instead proceeded much more rapidly at early times.
For instance, at fixed M$_{\ast}$, the average change in metallicity as a function of time would be roughly 0.01~dex~Gyr$^{-1}$ between $z=0$ and $z=1$, 0.03~dex~Gyr$^{-1}$ between $z=1$ and $z=2$, and 0.08~dex~Gyr$^{-1}$ between $z=2$ and $z=3$.
This accelerated evolution at early times is qualitatively consistent with the elevated SFRs observed at fixed M$_{\ast}$ in the early universe, as traced by the evolving star-forming main sequence \citep[e.g.,][]{2014ApJ...795..104W,2014ApJS..214...15S,2023MNRAS.519.1526P,2024ApJ...977..133C}. Higher SFRs imply more rapid metal production, though this enrichment must be balanced against metal dilution from gas accretion and metal removal from outflows. Thus, the faster metallicity evolution at high redshifts inferred from the MZR is likely driven by the interplay between vigorous star formation and gas flows at high redshift.

In our evolving MZR functional form (eq.~\ref{eq:2023_MZR_z}), the normalization evolution is parameterized in the redshift-dependent turnover mass above which the MZR flattens (M$_0$), which scales as approximately M$_0\propto (1+z)^{2.5}$.
Part of this evolution can be attributed to increasing gas fractions with redshift as $f_{\text{gas}}\propto (1+z)^{1.5}$ out to $z\sim3$ \citep[e.g.,][]{2017ApJ...837..150S,2020ARA&A..58..157T}. Higher gas fractions at earlier times lead to lower metallicities, contributing to the redshift evolution in MZR normalization.
\citet{2021ApJ...914...19S} found a similar rate of normalization decrease with increasing redshift as in this work, though with just three redshift bins.
These authors applied gas-regulator models to show that evolving $f_{\text{gas}}$ at fixed M$_{\ast}$ only accounted for roughly half of the evolution between $z=0$ and $z\sim3$, and attributed the remainder to an 
increase in $\zeta_\text{out}$ with increasing redshift such that outflows remain the dominant contributor governing the low-mass slope.
Early {\it JWST} results suggest that metallicity evolution at fixed $\rm M_*$ slows down at $z>4$, potentially due to a declining rate of increase in $f_{\text{gas}}$ as redshift increases \citep{2024A&A...684A..75C,2023ApJS..269...33N,2025ApJ...978..136S}.
However, a more complete understanding of the drivers of MZR normalization evolution requires a proper treatment of higher-order effects that are typically neglected such as inflowing metals, outflow recycling, departures from equilibrium, and transitions to bursty star-formation modes at high redshift and low mass \citep[e.g.,][]{2025arXiv250522712M}.
Our new results provide tight constraints to confront with theory.

\subsection{Comparison with simulations}
Observational constraints on MZR evolution provide a valuable benchmark for theoretical predictions from galaxy formation simulations. The colored lines in Figure~\ref{fig:Z10_gamma_evolution} display trends of star-forming galaxies in four simulations: IllustrisTNG \citep{2018MNRAS.473.4077P,2019MNRAS.484.5587T}, Illustris \citep{2014MNRAS.444.1518V,2014MNRAS.445..175G}, SIMBA \citep{2019MNRAS.486.2827D}, and EAGLE \citep{2015MNRAS.450.1937C,2015MNRAS.446..521S}. We used the median gas-phase oxygen abundance as a function of stellar mass extracted from these simulations and presented in \citet{2025MNRAS.536..119G}, limited to masses above $10^9\ \text{M}_\odot$. With these values taken from simulations, we fit the low-mass slope ($\gamma$) and metallicity at $10^{10}\ \text{M}_\odot$ using the same methods applied to the observational samples. These four simulations significantly diverge in their predictions of MZR slope and normalization due to differences in their implementations of feedback, star formation, stellar yields, and other baryonic processes \citep[for a detailed discussion, see][]{2024MNRAS.531.1398G,2025MNRAS.536..119G}.

The dashed lines in the left panel of Figure~\ref{fig:Z10_gamma_evolution} show the redshift evolution of the O/H at $10^{10}\ \text{M}_\odot$ from the simulations renormalized to match the observational data at $z=1.5$, allowing us to compare the relative metallicity evolution at fixed mass.
While all 4 simulations differ in absolute normalization by $\gtrsim0.1$~dex from the observations and one another, their relative metallicity evolution is much more similar and broadly captures our constraints on MZR evolution out to $z\sim3.3$.
IllustrisTNG and Illustris provide the closest match to our observational constraints.
SIMBA and EAGLE all predict a steeper metallicity evolution from $z=0-1.5$ than we find, while their rate of evolution declines at $z>1.5$ to better match our constraints, though SIMBA predicts that MZR normalization stops evolving above $z\sim2.5$ in disagreement with our results.
In SIMBA, the direction of MZR evolution reverses at $z\gtrsim4$, in conflict with {\it JWST} results at higher redshifts \citep{2023ApJS..269...33N, 2023ApJ...950L...1S,2024A&A...684A..75C}.
In IllustrisTNG, \citet{2019MNRAS.484.5587T} found that the downward evolution of MZR normalization was primarily driven by the significant increase in gas fraction at fixed mass with increasing redshift leading to greater dilution of ISM metals, and that this effect dominates over any changes in metal retention efficiency at fixed mass.

The right panel of Figure~\ref{fig:Z10_gamma_evolution} compares the redshift evolution of the low-mass MZR slope in the simulations to the observational constraints.
IllustrisTNG and SIMBA both display MZR slopes that do not strongly evolve, in qualitative agreement with our findings. In contrast, slopes in EAGLE strongly increase with redshift while those in Illustris decrease with redshift until $z\sim3$. IllustrisTNG is the closest quantitative match to the observational constraints, with $\gamma_\mathrm{TNG}\approx0.2-0.3$ over $z=0-4$.
 In contrast, SIMBA shows steeper slopes ($\gamma_\mathrm{SIMBA}\approx0.4$) than we find across most of this redshift range. Thus, while all four simulations broadly agree on the relative evolution of MZR normalization within $\sim0.1$~dex, they exhibit stark differences in the slope evolution.
Accordingly, the low-mass slope is a more powerful diagnostic than MZR normalization to distinguish between the predictions from this set of simulations. A non-evolving MZR slope of $\gamma\approx0.35$ is also seen in the FIRE simulations \citep{2016MNRAS.456.2140M,2024ApJ...967L..41M}, in good agreement with our constraints.
Extending this work to higher redshifts by applying our uniform analysis framework to representative samples at $z>4$ as they emerge from {\it JWST} observations will further test such simulations, which vary widely in their MZR predictions at $z\sim4-10$ \citep[e.g.,][]{2020MNRAS.494.1988L,2023MNRAS.518.3557U}.
In particular, determining whether the MZR slope evolves can reveal if the scaling of outflow loading factors with mass changes at high redshift, while normalization shifts can probe changes in gas fractions and metal retention.
Investigations of the underlying physical drivers of differences in gas fractions and metal retention efficiencies among simulations could provide further insight.

\subsection{Evolution of the FMR}
Early studies of the FMR at high redshifts indicated that the mean relation among stellar mass, SFR, and metallicity did not display much evolution over $z=0-2$, but its behavior beyond $z>2$ was debated. For instance, while \citet{2010MNRAS.408.2115M}  reported no significant evolution out to $z\sim2.3$ but a $-0.5$~dex offset in O/H at $z\sim3.5$, \citet{2010A&A...521L..53L} found no evolution up to $z\sim3.5$. \citep[see also][]{troncoso2014metallicity,2014MNRAS.440.2300C}.
Large offsets ($\gtrsim0.2$~dex) of $z\sim3$ samples from the $z\sim0$ FMR has largely disappeared in more recent works \citep[e.g.,][]{2018ApJ...858...99S,2021ApJ...914...19S,2019A&A...627A..42C,2020MNRAS.491..944C,2025ApJ...984..188K}, where the previously seen deviation is now understood to have arisen from biases in early NIR spectroscopic samples and a failure to account for evolving ISM ionization conditions when converting strong-line ratios to metallicity \citep[e.g.,][]{2021ApJ...914...19S}. Similar to these recent studies, we find that the mass-binned sample averages agree with the FMR observed at $z\sim0$ to better than $0.1$~dex out to $z\sim3.3$ (Fig.~\ref{fig:Residual} and Tab.~\ref{tab:Scatter_RMS}) regardless of the metallicity indicator used or the FMR parameterization adopted. The agreement with the \citet{2021ApJ...914...19S} $z\sim0$ FMR is particularly good, with nearly every mass bin at all redshifts offset by $\le0.05$~dex, likely stemming from the similarity of our metallicity estimation techniques.
Early work at higher redshifts with {\it JWST} has indicated potential evolution away from the FMR at $z>4$, but with contrasting results about the degree to which samples deviate from the $z\sim0$ FMR and at what redshift such evolution begins \citep[e.g.,][]{2024A&A...684A..75C, 2023arXiv230706336L, 2023ApJS..269...33N}.

A non-evolving FMR with relatively small scatter has been interpreted to indicate that galaxies oscillate about an equilibrium point at which gas accretion rates are roughly equal to the sum of SFRs and gas outflow rates such that the gas fraction does not change sharply with time, and that departures from equilibrium are relatively small and quickly dampened \citep[e.g.,][]{2010MNRAS.408.2115M,2013ApJ...772..119L,2012MNRAS.421...98D,2018MNRAS.477L..16T,2019MNRAS.484.5587T}.
For this interpretation to hold true, in addition to a non-evolution of the mean SFR-M$_{\ast}$-O/H relation, an anti-correlation between O/H and SFR at fixed M$_{\ast}$ must hold independently at each redshift with similar strength to that at $z\sim0$.
The mass-binned data employed in this study cannot be used to directly measure the strength of this secondary dependence.
\citet{2018ApJ...858...99S} found that a sample of 260 star-forming galaxies at $z\sim2.3$ from the MOSDEF survey indeed displayed an internal SFR-O/H anti-correlation at fixed mass with a similar slope to what is seen locally.
However, \citet{2025ApJ...984..188K} recently analyzed a similarly-sized sample at $z\sim2.3$ from the KBSS-MOSFIRE survey and did not find consistency with the $z\sim0$ FMR expectations internally, despite this sample also showing no offset from the $z\sim0$ relation on average.
These authors interpreted this result as an indication that most galaxies are not near equilibrium at this redshift. 
These findings also raise interesting questions, such as whether the FMR can truly persist out to high redshift, and if so, how can this be reconciled with models in which galaxies experience highly bursty star formation histories. Rapid fluctuations in SFR make it difficult for metallicity to remain tightly anti-correlated with SFR at fixed mass, thereby disrupting the coherence of the FMR.
Whether a tight FMR persists at early epochs could thus place strong constraints on the allowed timescales of SFR variability and gas processing in galaxy formation models. However, further work is needed to gain a clear picture of the true extent of FMR non-evolution and its physical implications.

An anti-correlation between SFR and metallicity at fixed mass is a nearly universal feature of analytic and numerical galaxy formation models at all redshifts \citep[e.g.,][]{2013ApJ...772..119L,2017MNRAS.472.3354D,2019MNRAS.486.2827D,2019MNRAS.484.5587T,2023MNRAS.518.3557U}.
The observational constraints presented here cannot accommodate more than $\approx0.1$~dex in evolutionary offset from the $z\sim0$ FMR out to $z\sim3.3$ at masses $>10^9\ \text{M}_\odot$.
An invariant or very weakly evolving FMR is present in the IllustrisTNG and Illustris simulations, but does not occur in EAGLE or SIMBA that already display deviations of $>0.3$~dex in O/H at fixed $\text{M}_\ast$ and SFR by $z=3$ \citep{2024MNRAS.531.1398G,2025MNRAS.536..119G}.
The degree of FMR evolution provides further evidence that the IllustrisTNG model performs best in reproducing our results.

\subsection{Areas of future improvement}

Throughout our analysis, we have attempted to mitigate systematic biases on the comparison of metallicities across redshift by using a uniform set of line ratios for all redshift and mass bins, and adopting different strong-line calibration sets at $z<1$ and $z>1$ to reflect evolving ISM ionization conditions.
However, there remain areas that could be further homogenized to increase the fidelity of future metallicity evolution studies.
While we have shifted all masses and SFRs onto a consistent \citet{2003PASP..115..763C} IMF scale, the techniques to estimate these properties vary among the samples.
We have taken the stellar masses as reported from the literature sources, except for the case of SDSS at $z\sim0.08$ where we shifted the masses down by 0.2~dex following the advice of \citet{2011ApJ...730..137Z} and \citet{2024ApJ...964...59L}.
A more uniform approach modeling the available photometry for each sample with the same techniques could further eliminate systematic biases between masses at each redshift, including attempts to understand how the more limited wavelength range sampled at some redshifts might impact the outcome.
Likewise, SFRs have also been adopted from the literature sources and could benefit from a uniform derivation for FMR analyses.

Finally, while the use of the \citet{2020MNRAS.491..944C} calibrations at $z<1$ and the \citet{2018ApJ...859..175B} analog calibrations at $z>1$ likely reduces systematic effects comparing metallicities of high- and low-redshift samples, this approach remains an indirect way of accounting for the ISM conditions present at high redshift.
The \citet{2018ApJ...859..175B} analog calibrations were selected to match a sample  of $z\sim2$ star-forming galaxies in the [OIII]/H$\beta$ vs.\ [NII]/H$\alpha$ line ratio diagram, but many unanswered questions remain about how closely the properties in that sample match actual high-redshift sources.
An obvious improvement would be to simply use strong-line calibrations that are constructed based on actual high-redshift samples.
The number of galaxies with temperature-sensitive auroral emission line detections from {\it JWST} is starting to become large enough for robust strong-line calibrations to be assessed in situ in the early universe \citep[e.g.,][]{2023ApJS..269...33N,2024ApJ...962...24S,2025arXiv250210499S,2024arXiv241215435C,2025arXiv250403839C}, but these samples still notably lack metal-rich galaxies (12+log(O/H$)>8.4$) in a range necessary for application to moderately massive galaxies at $z\sim1.5-3$.
Ultimately, a calibration that accounts for evolving ISM conditions as a continuous function of redshift is desirable, in lieu of changing calibrations above and below a single threshold redshift.
As the number of high-redshift galaxies with direct-method metallicities expands, these results can be reassessed with updated high-redshift calibrations.

\section{SUMMARY AND CONCLUSIONS}\label{sec:Conclusions}

We performed a uniform analysis of the evolution of the mass-metallicity relation (MZR) and fundamental metallicity relation (FMR) using samples in 6 redshift bins spanning $z=0-3.3$ and log(M$_{\ast}$/M$_\odot)=9.0-11.0$, achieving the finest time sampling ($1-3$~Gyr) to-date over a 12~Gyr baseline.
Metallicities in all 6 samples were derived from the same combination of the [OII], H$\beta$, and [OIII] rest-optical emission lines that allow use of the O3, O32, and R23 ratios, enabled at $z\sim1.5$ by new DEIMOS spectroscopy covering [OII] presented here.
We also derived metallicities from the commonly employed, but nitrogen-sensitive, O3N2 and N2 line ratios for comparison.
The impact of evolving ISM ionization conditions on the translation between strong-line ratio and O/H was accounted for by adopting a different set of calibrations for samples at $z<1$ \citep{2020MNRAS.491..944C} and $z>1$ \citep{2018ApJ...859..175B}, both based on empirical $T_e$ measurements.
We summarize the main results of the paper as follows:
\begin{itemize}
    \item At fixed redshift, rest-optical emission-line ratios uniformly show trends as a function of stellar mass indicating the existence of a positive correlation between O/H and M$_\ast$ in each redshift bin (Fig.~\ref{fig:Lines_ratios}). At fixed mass, we find monotonic evolution in line ratios indicating decreasing metallicity with increasing redshift. Samples at $z<1$ display flattening in the line ratios at high masses ($\gtrsim10^{10}\ \text{M}_\odot$).
    \item When the same metallicity indicator is employed with redshift-appropriate calibrations, we find strong evidence that the low-mass power law slope of the MZR does not evolve from $z\sim0$ to $z\sim3.3$ (Figs.~\ref{fig:MZR_all_3}, \ref{fig:MZR_all_z}, and~\ref{fig:Z10_gamma_evolution}; Tabs.~\ref{tab:BF-MZR_O3O2}$-$\ref{tab:BF-MZR_N2}). We find that, at low masses, metallicity scales approximately as $\text{O/H}\propto \rm M_*^{0.3}$, with only slight variations in steepness when employing different metallicity indicators. The non-evolution of this slope suggests that the same physical mechanism, namely metal removal via enriched outflows, controls the shape of the MZR and operates in moderately massive galaxies over the past 12~Gyr of cosmic time. Accordingly, the outflow metal loading factor is implied to scale as $\zeta_\text{out}\propto \rm M_*^{-0.3}$ without significant evolution out to at least $z\sim3.3$. A joint fit to all redshift bins with a redshift-invariant low-mass slope of $\gamma=0.28\pm0.01$ can simultaneously match observations in 6 redshift bins (Fig.~\ref{fig:MZR_all_z}; eq.~\ref{eq:2023_MZR_z}).
    \item We find that the normalization of the MZR at a fixed mass of $10^{10}$~M$_\odot$ evolves downward with increasing redshift at a rate that is constant with redshift as $d\log(\text{O/H})/dz =-0.11\pm0.01$ (Fig.~\ref{fig:Z10_gamma_evolution}). Accordingly, at fixed stellar mass, enrichment proceeded much more rapidly as a function of time at high redshifts than at $z\sim0$, reflecting higher SFRs and metal production rates at fixed M$_\ast$.
    \item We find no evidence that the high-mass asymptotic metallicity evolves with redshift. The samples at $z\sim0.08$, 0.3, and 0.8 are well-sampled through the turnover and clearly converge in metallicity at high masses ($\sim 10^{10.5}\ \text{M}_\odot$). At $z\ge1.5$, both sample size and possible incompleteness at high masses prevent any robust constraints on MZR flattening.
    \item The comparison of the measured metallicities in all 6 redshift bins to the best-fit $z\sim0$ FMR relations from \citet{2021ApJ...914...19S} and \citet{2020MNRAS.491..944C} reveals minimal FMR evolution out to $z\sim3.3$ (Figs.~\ref{fig:Curti_comparison} and \ref{fig:Residual}). The median offset in each redshift bin is $\Delta \log(\text{O/H})\le 0.1$~dex. This weak or absent FMR evolution suggests the possibility that galaxies remain near an equilibrium state up to $z\sim3.3$, but future work is required to confirm whether SFR$-$O/H anti-correlations at fixed mass exist internal to each redshift bin as expected in such a scenario.
    \item We compare our observational constraints on MZR and FMR evolution with four different cosmological simulations of galaxy formation (IllustrisTNG, Illustris, SIMBA, and EAGLE), and find that IllustrisTNG is the only one of these that simultaneously matches the non-evolving MZR slope, MZR normalization decrease, and weak or absent FMR evolution.
\end{itemize}

This work provides the most continuous, detailed, and uniform view of metallicity evolution over $z=0-4$ achieved to-date, setting a new benchmark for the MZR and FMR spanning from the present day through Cosmic Noon and defining the baseline for the exploration of metallicity evolution at higher redshifts with {\it JWST}.

\appendix
\twocolumngrid
\restartappendixnumbering
\section{Consistency between MOSFIRE and DEIMOS [OII] measurements}\label{app:stacking}
\begin{figure}[h!]
\includegraphics[width=1.\columnwidth]{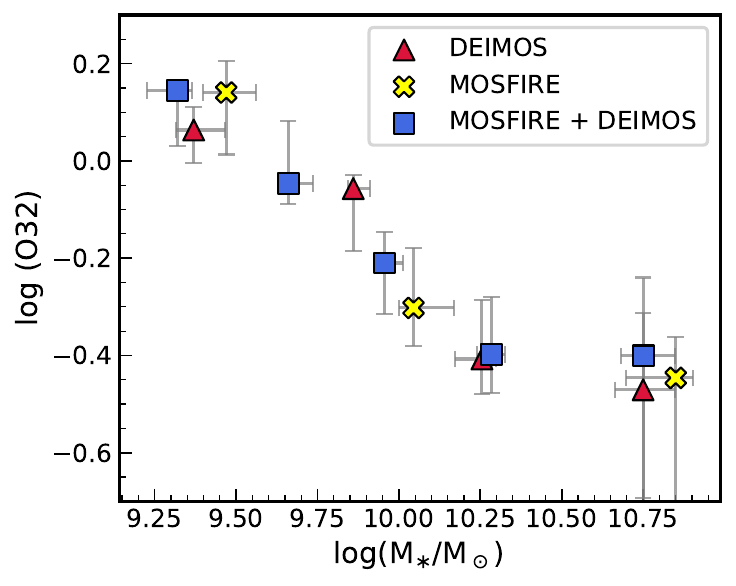}
\caption{O32 as a function of M$_{\ast}$ for mass-binned composite spectra of the $z\sim1.5$ star-forming galaxy sample.
Separate stacks were made including galaxies for which [OII] was covered with MOSFIRE (yellow crosses) and DEIMOS (red triangles).
The blue squares show our fiducial $z\sim1.5$ stacks in which [OII] spectra from both DEIMOS and MOSFIRE are included. The close agreement between the DEIMOS and MOSFIRE subsets demonstrates that the flux calibration is accurate to within the statistical uncertainty and that combining spectra from the two instruments does not introduce significant bias.
\label{fig:DEIMOS_vs_MOSFIRE}} 
\end{figure}

To test the consistency between spectra from DEIMOS and MOSFIRE covering [OII], we compared O32 ratios measured from mass-binned composites constructed for two subsamples, one containing galaxies with [OII] observations from DEIMOS and the other with [OII] measurements from MOSFIRE. Figure~\ref{fig:DEIMOS_vs_MOSFIRE} compares the composite spectra of these subsamples to the fiducial composites based on the combined sample that is used in our analysis. The MOSFIRE-[OII] stacks (yellow crosses) include 42 galaxies divided into three mass bins. The DEIMOS-[OII] (red triangles) sample consists of 86 galaxies, stacked into four mass bins. Because DEIMOS spectra have roughly twice the spectral resolution of MOSFIRE, they were smoothed and resampled to match the MOSFIRE $Y$ band resolution before stacking. Thus, the DEIMOS stacks shown are based on these smoothed spectra. The fiducial MOSFIRE+DEIMOS stacks used in our analysis are shown as blue squares. The relations between O32 and M$_\ast$ displayed by the two subsets are fully consistent with one another within the uncertainties, confirming that no significant biases are introduced by combining the DEIMOS and MOSFIRE [OII] spectra when producing our fiducial $z\sim1.5$ composites, and indicating that the flux calibration is accurate to within the statistical uncertainty.

\section{Properties of the mass-binned literature samples}\label{app:literature}
Table~\ref{tab:Properties_all_z} summarizes the properties of mass-binned composites for the five literature samples included in this analysis. The columns list the median stellar mass, SFR, rest-optical emission-line ratios, and the corresponding gas-phase metallicities inferred from each metallicity indicator (indicated by superscripts).

\setlength{\tabcolsep}{4pt}
\startlongtable
\begin{deluxetable*}{lcccccccccc}
\tabletypesize{\scriptsize}
\tablewidth{0pt}
\tablecaption{Properties of the mass-binned literature samples\label{tab:Properties_all_z}}
\tabletypesize{\tiny}
\tablehead{
\colhead{$\log \left(\frac{\text{M}_{\ast}}{\text{M}_{\odot}}\right)_{\text{med}}$} & \colhead{$\log \left(\frac{\text{SFR}}{\text{M}_{\odot}/\text{yr}}\right)_{\text{med}}$} & \colhead{$\log (\text{O3})$} & \colhead{$\log (\text{O2})$} & \colhead{$\log (\text{O32})$} & \colhead{$\log (\text{R23})$} & \colhead{$\log (\text{N2})$} & \colhead{$\log (\text{O3N2})$} & \colhead{$12 + \log \left(\frac{\text{O}}{\text{H}}\right)^{\text{O-based}}$} &
\colhead{$12 + \log \left(\frac{\text{O}}{\text{H}}\right)^{\text{O3N2}}$} &
\colhead{$12 + \log \left(\frac{\text{O}}{\text{H}}\right)^{\text{N2}}$}}
\colnumbers
\startdata
\multicolumn{11}{c}{\textbf{$z\sim0.08$}} \\
\hline
\hline
9.03$^{+0.05}_{-0.05}$ & -0.35$^{+0.25}_{-0.25}$ & 0.09$^{+0.01}_{-0.01}$ & 0.56$^{+0.01}_{-0.01}$ & -0.47$^{+0.01}_{-0.01}$ & 0.73$^{+0.01}_{-0.01}$ & -0.68$^{+0.01}_{-0.01}$ & 0.77$^{+0.01}_{-0.01}$ & 8.58$^{+0.01}_{-0.01}$ & 8.58$^{+0.01}_{-0.01}$ & 8.58$^{+0.01}_{-0.01}$ \\
9.12$^{+0.05}_{-0.05}$ & -0.26$^{+0.25}_{-0.25}$ & 0.05$^{+0.01}_{-0.01}$ & 0.56$^{+0.01}_{-0.01}$ & -0.51$^{+0.01}_{-0.01}$ & 0.71$^{+0.01}_{-0.01}$ & -0.65$^{+0.01}_{-0.01}$ & 0.70$^{+0.01}_{-0.01}$ & 8.59$^{+0.01}_{-0.01}$ & 8.60$^{+0.01}_{-0.01}$ & 8.60$^{+0.01}_{-0.01}$\\
9.23$^{+0.05}_{-0.05}$ & -0.21$^{+0.25}_{-0.25}$ & -0.02$^{+0.01}_{-0.01}$ & 0.54$^{+0.04}_{-0.05}$ & -0.55$^{+0.01}_{-0.01}$ & 0.68$^{+0.01}_{-0.01}$ & -0.62$^{+0.01}_{-0.01}$ & 0.60$^{+0.01}_{-0.01}$ & 8.62$^{+0.01}_{-0.01}$ & 8.62$^{+0.01}_{-0.01}$ & 8.61$^{+0.01}_{-0.01}$\\
9.33$^{+0.05}_{-0.05}$ & -0.14$^{+0.25}_{-0.25}$ & -0.09$^{+0.01}_{-0.01}$ & 0.52$^{+0.01}_{-0.01}$ & -0.60$^{+0.01}_{-0.01}$ & 0.64$^{+0.01}_{-0.01}$ & -0.58$^{+0.01}_{-0.01}$ & 0.50$^{+0.01}_{-0.01}$ & 8.64$^{+0.01}_{-0.01}$ & 8.64$^{+0.01}_{-0.01}$ & 8.63$^{+0.01}_{-0.01}$\\
9.43$^{+0.05}_{-0.05}$ & -0.08$^{+0.25}_{-0.25}$ & -0.15$^{+0.01}_{-0.01}$ & 0.50$^{+0.01}_{-0.01}$ & -0.64$^{+0.01}_{-0.01}$ & 0.61$^{+0.01}_{-0.01}$ & -0.55$^{+0.01}_{-0.01}$ & 0.41$^{+0.01}_{-0.01}$ & 8.65$^{+0.01}_{-0.01}$ & 8.66$^{+0.01}_{-0.01}$ & 8.65$^{+0.01}_{-0.01}$\\
9.53$^{+0.05}_{-0.05}$ & -0.04$^{+0.25}_{-0.25}$ & -0.21$^{+0.01}_{-0.01}$ & 0.47$^{+0.01}_{-0.01}$ & -0.68$^{+0.01}_{-0.01}$ & 0.58$^{+0.01}_{-0.01}$ & -0.52$^{+0.01}_{-0.01}$ & 0.32$^{+0.01}_{-0.01}$ & 8.67$^{+0.01}_{-0.01}$ & 8.68$^{+0.01}_{-0.01}$ & 8.67$^{+0.01}_{-0.01}$\\
9.63$^{+0.05}_{-0.05}$ & 0.04$^{+0.25}_{-0.25}$ & -0.24$^{+0.01}_{-0.01}$ & 0.45$^{+0.01}_{-0.01}$ & -0.69$^{+0.01}_{-0.01}$ & 0.56$^{+0.01}_{-0.01}$ & -0.51$^{+0.01}_{-0.01}$ & 0.27$^{+0.01}_{-0.01}$ & 8.68$^{+0.01}_{-0.01}$ & 8.69$^{+0.01}_{-0.01}$ & 8.68$^{+0.01}_{-0.01}$\\
9.73$^{+0.05}_{-0.05}$ & 0.09$^{+0.25}_{-0.25}$ & -0.30$^{+0.01}_{-0.01}$ & 0.43$^{+0.01}_{-0.01}$ & -0.72$^{+0.01}_{-0.01}$ & 0.52$^{+0.01}_{-0.01}$ & -0.49$^{+0.01}_{-0.01}$ & 0.19$^{+0.01}_{-0.01}$ & 8.70$^{+0.01}_{-0.01}$ & 8.71$^{+0.01}_{-0.01}$ & 8.69$^{+0.01}_{-0.01}$\\
9.83$^{+0.05}_{-0.05}$ & 0.15$^{+0.25}_{-0.25}$ & -0.35$^{+0.01}_{-0.01}$ & 0.40$^{+0.01}_{-0.01}$ & -0.75$^{+0.01}_{-0.01}$ & 0.49$^{+0.01}_{-0.01}$ & -0.46$^{+0.01}_{-0.01}$ & 0.11$^{+0.01}_{-0.01}$ & 8.71$^{+0.01}_{-0.01}$ & 8.73$^{+0.01}_{-0.01}$ & 8.71$^{+0.01}_{-0.01}$\\
9.93$^{+0.05}_{-0.05}$ & 0.22$^{+0.25}_{-0.25}$ & -0.39$^{+0.01}_{-0.01}$ & 0.37$^{+0.01}_{-0.01}$ & -0.76$^{+0.01}_{-0.01}$ & 0.46$^{+0.01}_{-0.01}$ & -0.45$^{+0.01}_{-0.01}$ & 0.05$^{+0.01}_{-0.01}$ & 8.73$^{+0.01}_{-0.01}$ & 8.74$^{+0.01}_{-0.01}$ & 8.72$^{+0.01}_{-0.01}$\\
10.03$^{+0.05}_{-0.05}$ & 0.37$^{+0.25}_{-0.25}$ & -0.43$^{+0.01}_{-0.01}$ & 0.34$^{+0.01}_{-0.01}$ & -0.77$^{+0.01}_{-0.01}$ & 0.43$^{+0.01}_{-0.01}$ & -0.44$^{+0.01}_{-0.01}$ & 0.01$^{+0.01}_{-0.01}$ & 8.74$^{+0.01}_{-0.01}$ & 8.75$^{+0.01}_{-0.01}$ & 8.73$^{+0.01}_{-0.01}$\\
10.13$^{+0.05}_{-0.05}$ & 0.43$^{+0.25}_{-0.25}$ & -0.45$^{+0.01}_{-0.01}$ & 0.32$^{+0.01}_{-0.01}$ & -0.78$^{+0.01}_{-0.01}$ & 0.41$^{+0.01}_{-0.01}$ & -0.43$^{+0.01}_{-0.01}$ & -0.03$^{+0.01}_{-0.01}$ & 8.74$^{+0.01}_{-0.01}$ & 8.75$^{+0.01}_{-0.01}$ & 8.73$^{+0.01}_{-0.01}$\\
10.23$^{+0.05}_{-0.05}$ & 0.57$^{+0.25}_{-0.25}$ & -0.47$^{+0.01}_{-0.01}$ & 0.32$^{+0.01}_{-0.01}$ & -0.78$^{+0.01}_{-0.01}$ & 0.40$^{+0.01}_{-0.01}$ & -0.42$^{+0.01}_{-0.01}$ & -0.05$^{+0.01}_{-0.01}$ & 8.75$^{+0.01}_{-0.01}$ & 8.76$^{+0.01}_{-0.01}$ & 8.74$^{+0.01}_{-0.01}$\\
10.33$^{+0.05}_{-0.05}$ & 0.53$^{+0.25}_{-0.25}$ & -0.47$^{+0.01}_{-0.01}$ & 0.30$^{+0.01}_{-0.01}$ & -0.77$^{+0.01}_{-0.01}$ & 0.39$^{+0.01}_{-0.01}$ & -0.41$^{+0.01}_{-0.01}$ & -0.06$^{+0.01}_{-0.01}$ & 8.75$^{+0.01}_{-0.01}$ & 8.76$^{+0.01}_{-0.01}$ & 8.75$^{+0.01}_{-0.01}$\\
10.43$^{+0.05}_{-0.05}$ & 0.60$^{+0.25}_{-0.25}$ & -0.47$^{+0.01}_{-0.01}$ & 0.29$^{+0.01}_{-0.01}$ & -0.77$^{+0.01}_{-0.01}$ & 0.38$^{+0.01}_{-0.01}$ & -0.40$^{+0.01}_{-0.01}$ & -0.07$^{+0.01}_{-0.01}$ & 8.75$^{+0.01}_{-0.01}$ & 8.76$^{+0.01}_{-0.01}$ & 8.76$^{+0.01}_{-0.01}$\\
10.53$^{+0.05}_{-0.05}$ & 0.74$^{+0.25}_{-0.25}$ & -0.48$^{+0.01}_{-0.01}$ & 0.29$^{+0.01}_{-0.01}$ & -0.77$^{+0.01}_{-0.01}$ & 0.38$^{+0.01}_{-0.01}$ & -0.40$^{+0.01}_{-0.01}$ & -0.08$^{+0.01}_{-0.01}$ & 8.75$^{+0.01}_{-0.01}$ & 8.76$^{+0.01}_{-0.01}$ & 8.76$^{+0.01}_{-0.01}$\\
10.63$^{+0.05}_{-0.05}$ & 0.83$^{+0.25}_{-0.25}$ & -0.47$^{+0.01}_{-0.01}$ & 0.29$^{+0.01}_{-0.01}$ & -0.76$^{+0.01}_{-0.01}$ & 0.38$^{+0.01}_{-0.01}$ & -0.39$^{+0.01}_{-0.01}$ & -0.08$^{+0.01}_{-0.01}$ & 8.75$^{+0.01}_{-0.01}$ & 8.76$^{+0.01}_{-0.01}$ & 8.77$^{+0.01}_{-0.01}$\\
10.73$^{+0.05}_{-0.05}$ & 0.91$^{+0.25}_{-0.25}$ & -0.45$^{+0.01}_{-0.01}$ & 0.29$^{+0.01}_{-0.01}$ & -0.74$^{+0.01}_{-0.01}$ & 0.38$^{+0.01}_{-0.01}$ & -0.38$^{+0.01}_{-0.01}$ & -0.07$^{+0.01}_{-0.01}$ & 8.75$^{+0.01}_{-0.01}$ & 8.76$^{+0.01}_{-0.01}$ & 8.77$^{+0.01}_{-0.01}$\\
\hline
\hline
\multicolumn{11}{c}{\textbf{$z\sim0.3$}} \\
\hline
\hline
9.19$^{+0.04}_{-0.05}$ & 0.10$^{+0.04}_{-0.05}$ & 0.23$^{+0.01}_{-0.01}$ & 0.57$^{+0.01}_{-0.01}$ & -0.33$^{+0.01}_{-0.01}$ & 0.78$^{+0.01}_{-0.01}$ & -0.72$^{+0.01}_{-0.01}$ & 0.95$^{+0.01}_{-0.01}$ & 8.53$^{+0.01}_{-0.01}$ & 8.54$^{+0.01}_{-0.01}$ & 8.56$^{+0.01}_{-0.01}$\\
9.39$^{+0.04}_{-0.05}$ & 0.27$^{+0.04}_{-0.05}$ & 0.09$^{+0.01}_{-0.01}$ & 0.54$^{+0.01}_{-0.01}$ & -0.45$^{+0.01}_{-0.01}$ & 0.71$^{+0.01}_{-0.01}$ & -0.61$^{+0.01}_{-0.01}$ & 0.70$^{+0.01}_{-0.01}$ & 8.58$^{+0.01}_{-0.01}$ & 8.60$^{+0.01}_{-0.01}$ & 8.62$^{+0.01}_{-0.01}$\\
9.54$^{+0.04}_{-0.05}$ & 0.30$^{+0.04}_{-0.05}$ & 0.00$^{+0.01}_{-0.01}$ & 0.56$^{+0.01}_{-0.01}$ & -0.56$^{+0.01}_{-0.01}$ & 0.69$^{+0.01}_{-0.01}$ & -0.56$^{+0.01}_{-0.01}$ & 0.56$^{+0.01}_{-0.01}$ & 8.61$^{+0.01}_{-0.01}$ & 8.63$^{+0.01}_{-0.01}$ & 8.64$^{+0.01}_{-0.01}$\\
9.69$^{+0.04}_{-0.05}$ & 0.35$^{+0.04}_{-0.05}$ & -0.07$^{+0.01}_{-0.01}$ & 0.53$^{+0.01}_{-0.01}$ & -0.59$^{+0.01}_{-0.01}$ & 0.65$^{+0.01}_{-0.01}$ & -0.53$^{+0.01}_{-0.01}$ & 0.47$^{+0.01}_{-0.01}$ & 8.63$^{+0.01}_{-0.01}$ & 8.65$^{+0.01}_{-0.01}$ & 8.66$^{+0.01}_{-0.01}$\\
9.80$^{+0.04}_{-0.05}$ & 0.42$^{+0.04}_{-0.05}$ & -0.16$^{+0.01}_{-0.01}$ & 0.52$^{+0.01}_{-0.01}$ & -0.68$^{+0.01}_{-0.01}$ & 0.63$^{+0.01}_{-0.01}$ & -0.48$^{+0.01}_{-0.01}$ & 0.32$^{+0.01}_{-0.01}$ & 8.66$^{+0.01}_{-0.01}$ & 8.68$^{+0.01}_{-0.01}$ & 8.69$^{+0.01}_{-0.01}$\\
9.94$^{+0.04}_{-0.05}$ & 0.55$^{+0.04}_{-0.05}$ & -0.17$^{+0.01}_{-0.01}$ & 0.51$^{+0.01}_{-0.01}$ & -0.68$^{+0.01}_{-0.01}$ & 0.61$^{+0.01}_{-0.01}$ & -0.49$^{+0.01}_{-0.01}$ & 0.32$^{+0.01}_{-0.01}$ & 8.67$^{+0.01}_{-0.01}$ & 8.68$^{+0.01}_{-0.01}$ & 8.69$^{+0.01}_{-0.01}$\\
10.06$^{+0.04}_{-0.05}$ & 0.59$^{+0.04}_{-0.05}$ & -0.29$^{+0.01}_{-0.01}$ & 0.48$^{+0.01}_{-0.01}$ & -0.76$^{+0.01}_{-0.01}$ & 0.57$^{+0.01}_{-0.01}$ & -0.44$^{+0.01}_{-0.01}$ & 0.15$^{+0.01}_{-0.01}$ & 8.70$^{+0.01}_{-0.01}$ & 8.72$^{+0.01}_{-0.01}$ & 8.72$^{+0.01}_{-0.01}$\\
10.21$^{+0.04}_{-0.05}$ & 0.69$^{+0.04}_{-0.05}$ & -0.39$^{+0.01}_{-0.01}$ & 0.45$^{+0.01}_{-0.01}$ & -0.84$^{+0.01}_{-0.01}$ & 0.53$^{+0.01}_{-0.01}$ & -0.41$^{+0.01}_{-0.01}$ & 0.02$^{+0.01}_{-0.01}$ & 8.72$^{+0.01}_{-0.01}$ & 8.74$^{+0.01}_{-0.01}$ & 8.75$^{+0.01}_{-0.01}$\\
10.40$^{+0.04}_{-0.05}$ & 0.83$^{+0.04}_{-0.05}$ & -0.37$^{+0.01}_{-0.01}$ & 0.48$^{+0.01}_{-0.01}$ & -0.85$^{+0.01}_{-0.01}$ & 0.55$^{+0.01}_{-0.01}$ & -0.40$^{+0.01}_{-0.01}$ & 0.03$^{+0.01}_{-0.01}$ & 8.72$^{+0.01}_{-0.01}$ & 8.74$^{+0.01}_{-0.01}$ & 8.75$^{+0.01}_{-0.01}$\\
10.67$^{+0.04}_{-0.05}$ & 0.88$^{+0.04}_{-0.05}$ & -0.47$^{+0.02}_{-0.02}$ & 0.42$^{+0.01}_{-0.01}$ & -0.90$^{+0.01}_{-0.01}$ & 0.49$^{+0.02}_{-0.02}$ & -0.36$^{+0.01}_{-0.01}$ & -0.11$^{+0.02}_{-0.02}$ & 8.74$^{+0.01}_{-0.01}$ & 8.77$^{+0.01}_{-0.01}$ & 8.80$^{+0.01}_{-0.01}$\\
\hline
\hline
\multicolumn{11}{c}{\textbf{$z\sim0.8$}} \\
\hline
\hline
9.19$^{+0.04}_{-0.05}$ & 0.37$^{+0.04}_{-0.05}$ & 0.40$^{+0.04}_{-0.05}$ & 0.50$^{+0.04}_{-0.05}$ & -0.10$^{+0.04}_{-0.05}$ & 0.81$^{+0.04}_{-0.05}$ & - & - & 8.46$^{+0.01}_{-0.01}$  & - & -\\
9.29$^{+0.04}_{-0.05}$ & 0.39$^{+0.04}_{-0.05}$ & 0.30$^{+0.04}_{-0.05}$ & 0.53$^{+0.04}_{-0.05}$ & -0.23$^{+0.04}_{-0.05}$ & 0.78$^{+0.04}_{-0.05}$ & - & - & 8.50$^{+0.01}_{-0.01}$ & - & -\\
9.37$^{+0.04}_{-0.05}$ & 0.43$^{+0.04}_{-0.05}$ & 0.28$^{+0.04}_{-0.05}$ & 0.51$^{+0.04}_{-0.05}$ & -0.23$^{+0.04}_{-0.05}$ & 0.76$^{+0.04}_{-0.05}$ & - & - & 8.51$^{+0.01}_{-0.01}$  & - & -\\
9.44$^{+0.04}_{-0.05}$ & 0.49$^{+0.04}_{-0.05}$ & 0.23$^{+0.04}_{-0.05}$ & 0.55$^{+0.04}_{-0.05}$ & -0.32$^{+0.04}_{-0.05}$ & 0.76$^{+0.04}_{-0.05}$ & - & - & 8.54$^{+0.01}_{-0.01}$  & - & -\\
9.51$^{+0.04}_{-0.05}$ & 0.54$^{+0.04}_{-0.05}$ & 0.24$^{+0.04}_{-0.05}$ & 0.57$^{+0.04}_{-0.05}$ & -0.32$^{+0.04}_{-0.05}$ & 0.78$^{+0.04}_{-0.05}$ & - & - & 8.53$^{+0.01}_{-0.01}$  & - & -\\
9.60$^{+0.04}_{-0.05}$ & 0.60$^{+0.04}_{-0.05}$ & 0.19$^{+0.04}_{-0.05}$ & 0.53$^{+0.04}_{-0.05}$ & -0.34$^{+0.04}_{-0.05}$ & 0.74$^{+0.04}_{-0.05}$ & - & - & 8.55$^{+0.01}_{-0.01}$  & - & -\\
9.70$^{+0.04}_{-0.05}$ & 0.60$^{+0.04}_{-0.05}$ & 0.11$^{+0.04}_{-0.05}$ & 0.52$^{+0.04}_{-0.05}$ & -0.41$^{+0.04}_{-0.05}$ & 0.70$^{+0.04}_{-0.05}$ & - & - & 8.57$^{+0.01}_{-0.01}$  & - & -\\
9.79$^{+0.04}_{-0.05}$ & 0.72$^{+0.04}_{-0.05}$ & 0.03$^{+0.04}_{-0.05}$ & 0.53$^{+0.04}_{-0.05}$ & -0.50$^{+0.04}_{-0.05}$ & 0.68$^{+0.04}_{-0.05}$ & - & - & 8.61$^{+0.01}_{-0.01}$  & - & -\\
9.87$^{+0.04}_{-0.05}$ & 0.77$^{+0.04}_{-0.05}$ & -0.02$^{+0.04}_{-0.05}$ & 0.48$^{+0.04}_{-0.05}$ & -0.50$^{+0.04}_{-0.05}$ & 0.63$^{+0.04}_{-0.05}$ & - & - & 8.62$^{+0.01}_{-0.01}$  & - & -\\
9.99$^{+0.04}_{-0.05}$ & 0.82$^{+0.04}_{-0.05}$ & -0.09$^{+0.04}_{-0.05}$ & 0.45$^{+0.04}_{-0.05}$ & -0.54$^{+0.04}_{-0.05}$ & 0.59$^{+0.04}_{-0.05}$ & - & - & 8.64$^{+0.01}_{-0.01}$  & - & -\\
10.12$^{+0.04}_{-0.05}$ & 0.92$^{+0.04}_{-0.05}$ & -0.12$^{+0.04}_{-0.05}$ & 0.47$^{+0.04}_{-0.05}$ & -0.59$^{+0.04}_{-0.05}$ & 0.60$^{+0.04}_{-0.05}$ & - & - & 8.65$^{+0.01}_{-0.01}$  & - & -\\
10.29$^{+0.04}_{-0.05}$ & 0.98$^{+0.04}_{-0.05}$ & -0.20$^{+0.04}_{-0.05}$ & 0.42$^{+0.04}_{-0.05}$ & -0.62$^{+0.04}_{-0.05}$ & 0.54$^{+0.04}_{-0.05}$ & - & - & 8.67$^{+0.01}_{-0.01}$  & - & -\\
10.56$^{+0.04}_{-0.05}$ & 1.29$^{+0.04}_{-0.05}$ & -0.35$^{+0.04}_{-0.05}$ & 0.33$^{+0.04}_{-0.05}$ & -0.67$^{+0.04}_{-0.05}$ & 0.44$^{+0.04}_{-0.05}$ & - & - & 8.70$^{+0.01}_{-0.01}$  & - & -\\
10.04$^{+0.04}_{-0.05}$ & 0.95$^{+0.04}_{-0.05}$ & -0.16$^{+0.04}_{-0.05}$ & 0.41$^{+0.04}_{-0.05}$ & -0.58$^{+0.04}_{-0.05}$ & 0.54$^{+0.04}_{-0.05}$ & - & - & 8.66$^{+0.01}_{-0.01}$  & - & -\\
10.22$^{+0.04}_{-0.05}$ & 1.07$^{+0.04}_{-0.05}$ & -0.29$^{+0.04}_{-0.05}$ & 0.38$^{+0.04}_{-0.05}$ & -0.68$^{+0.04}_{-0.05}$ & 0.49$^{+0.04}_{-0.05}$ & - & - & 8.70$^{+0.01}_{-0.01}$  & - & -\\
10.38$^{+0.04}_{-0.05}$ & 1.23$^{+0.04}_{-0.05}$ & -0.32$^{+0.04}_{-0.05}$ & 0.35$^{+0.04}_{-0.05}$ & -0.68$^{+0.04}_{-0.05}$ & 0.46$^{+0.04}_{-0.05}$ & - & - & 8.70$^{+0.01}_{-0.01}$  & - & -\\
10.54$^{+0.04}_{-0.05}$ & 1.30$^{+0.04}_{-0.05}$ & -0.40$^{+0.04}_{-0.05}$ & 0.28$^{+0.04}_{-0.05}$ & -0.68$^{+0.04}_{-0.05}$ & 0.39$^{+0.04}_{-0.05}$ & - & - & 8.73$^{+0.01}_{-0.01}$  & - & -\\
10.69$^{+0.04}_{-0.05}$ & 1.34$^{+0.04}_{-0.05}$ & -0.43$^{+0.04}_{-0.05}$ & 0.30$^{+0.04}_{-0.05}$ & -0.73$^{+0.04}_{-0.05}$ & 0.40$^{+0.04}_{-0.05}$ & - & - & 8.73$^{+0.01}_{-0.01}$  & - & -\\
10.82$^{+0.04}_{-0.05}$ & 1.37$^{+0.04}_{-0.05}$ & -0.46$^{+0.04}_{-0.05}$ & 0.23$^{+0.04}_{-0.05}$ & -0.68$^{+0.04}_{-0.05}$ & 0.33$^{+0.04}_{-0.05}$ & - & - & 8.74$^{+0.01}_{-0.01}$  & - & -\\
\hline
\hline
\multicolumn{11}{c}{\textbf{$z\sim2.3$}} \\
\hline
\hline
9.33$^{+0.01}_{-0.05}$ & 1.00$^{+0.13}_{-0.01}$ & 0.68$^{+0.01}_{-0.04}$ & 0.47$^{+0.05}_{-0.03}$ & 0.21$^{+0.01}_{-0.08}$ & 0.97$^{+0.01}_{-0.03}$ & -1.22$^{+0.08}_{-0.05}$ & 1.90$^{+0.05}_{-0.12}$ & 8.27$^{+0.02}_{-0.02}$ & 8.29$^{+0.04}_{-0.03}$ & 8.34$^{+0.04}_{-0.03}$\\
9.62$^{+0.03}_{-0.01}$ & 1.06$^{+0.11}_{-0.05}$ & 0.56$^{+0.04}_{-0.01}$ & 0.51$^{+0.05}_{-0.03}$ & 0.05$^{+0.04}_{-0.04}$ & 0.91$^{+0.04}_{-0.02}$ & -1.11$^{+0.07}_{-0.06}$ & 1.67$^{+0.08}_{-0.08}$ & 8.39$^{+0.02}_{-0.01}$ & 8.38$^{+0.03}_{-0.03}$ & 8.40$^{+0.03}_{-0.03}$\\
9.89$^{+0.02}_{-0.02}$ & 1.30$^{+0.11}_{-0.03}$ & 0.53$^{+0.03}_{-0.04}$ & 0.63$^{+0.03}_{-0.04}$ & -0.11$^{+0.04}_{-0.04}$ & 0.95$^{+0.02}_{-0.03}$ & -0.93$^{+0.04}_{-0.05}$ & 1.46$^{+0.06}_{-0.07}$ & 8.42$^{+0.02}_{-0.02}$ & 8.46$^{+0.03}_{-0.03}$ & 8.49$^{+0.02}_{-0.02}$\\
10.23$^{+0.02}_{-0.02}$ & 1.65$^{+0.05}_{-0.07}$ & 0.39$^{+0.03}_{-0.02}$ & 0.64$^{+0.01}_{-0.05}$ & -0.25$^{+0.05}_{-0.02}$ & 0.88$^{+0.02}_{-0.04}$ & -0.74$^{+0.04}_{-0.03}$ & 1.13$^{+0.04}_{-0.05}$ & 8.53$^{+0.01}_{-0.01}$ & 8.59$^{+0.02}_{-0.02}$ & 8.58$^{+0.02}_{-0.02}$\\
10.64$^{+0.03}_{-0.06}$ & 1.87$^{+0.09}_{-0.10}$ & 0.26$^{+0.07}_{-0.06}$ & 0.62$^{+0.10}_{-0.06}$ & -0.36$^{+0.10}_{-0.11}$ & 0.82$^{+0.07}_{-0.04}$ & -0.63$^{+0.08}_{-0.01}$ & 0.89$^{+0.06}_{-0.13}$ & 8.59$^{+0.03}_{-0.03}$ & 8.68$^{+0.04}_{-0.04}$ & 8.63$^{+0.02}_{-0.02}$\\
\hline
\hline
\multicolumn{11}{c}{\textbf{$z\sim3.3$}} \\
\hline
\hline
9.23$^{+0.02}_{-0.09}$ & 1.27$^{+0.14}_{-0.06}$ & 0.74$^{+0.05}_{-0.04}$ & 0.29$^{+0.11}_{-0.04}$ & 0.45$^{+0.06}_{-0.10}$ & 0.97$^{+0.04}_{-0.02}$ & - & - & 8.16$^{+0.04}_{-0.03}$ &  - & -\\
9.53$^{+0.02}_{-0.05}$ & 1.24$^{+0.16}_{-0.02}$ & 0.69$^{+0.06}_{-0.02}$ & 0.44$^{+0.08}_{-0.05}$ & 0.25$^{+0.06}_{-0.05}$ & 0.97$^{+0.06}_{-0.03}$ & - & - & 8.27$^{+0.03}_{-0.02}$ & - & -\\
9.82$^{+0.05}_{-0.03}$ & 1.57$^{+0.15}_{-0.08}$ & 0.65$^{+0.03}_{-0.07}$ & 0.50$^{+0.07}_{-0.05}$ & 0.15$^{+0.03}_{-0.09}$ & 0.96$^{+0.03}_{-0.05}$ & - & - & 8.32$^{+0.03}_{-0.03}$ & - & -\\
10.21$^{+0.04}_{-0.05}$ & 1.75$^{+0.09}_{-0.10}$ & 0.53$^{+0.07}_{-0.03}$ & 0.65$^{+0.05}_{-0.07}$ & -0.13$^{+0.09}_{-0.03}$ & 0.95$^{+0.05}_{-0.05}$ & - & - & 8.45$^{+0.02}_{-0.02}$ & - & -\\
10.60$^{+0.13}_{-0.01}$ & 2.26$^{+0.21}_{-0.28}$ & 0.35$^{+0.20}_{-0.14}$ & 0.62$^{+0.30}_{-0.16}$ & -0.27$^{+0.13}_{-0.17}$ & 0.87$^{+0.25}_{-0.11}$ & - & - & 8.56$^{+0.07}_{-0.06}$ & - & -\\
\enddata
\end{deluxetable*}


\bibliography{Bibliography}{}
\bibliographystyle{aasjournal}

\end{document}